\documentclass[prb,amssymb,twocolumn,showpacs,superscriptaddress,floatfix,a4paper]{revtex4}

\usepackage{graphicx}
\usepackage{bm}
\usepackage{amsmath}
\usepackage{mathptmx}

\begin{document}

\title{Langevin simulations of the out--of--equilibrium
dynamics of the vortex glass in high-temperature superconductors}
\author{Sebastian Bustingorry}
\affiliation{Centro At{\'{o}}mico Bariloche,
8400 San Carlos de Bariloche,
R\'{\i}o Negro, Argentina}
\author{Leticia F. Cugliandolo}
\affiliation{
Universit\'e Pierre et Marie Curie -- Paris VI, LPTHE UMR 7589,
4 Place Jussieu,  75252 Paris Cedex 05, France \\
and LPT Ecole Normale Sup{\'e}rieure, 24 rue Lhomond, 75231
Paris Cedex 05 France}
\author{Daniel Dom\'{\i}nguez}
\affiliation{Centro At{\'{o}}mico Bariloche,
8400 San Carlos de Bariloche,
R\'{\i}o Negro, Argentina}
\date{\today}

\begin{abstract}

We study the relaxation dynamics of flux lines in dirty
high-temperature superconductors using numerical simulations of a
London-Langevin model of the interacting vortex lines. By analysing
the equilibrium dynamics in the vortex liquid phase we find a dynamic
crossover to a glassy non-equilibrium regime. We then focus on the
out-of-equilibrium dynamics of the vortex glass phase using tools that
are common in the study of other glassy systems. By monitoring the
two-times roughness and dynamic wandering we identify and
characterize finite-size effects that are similar, though more
complex, than the ones found in the stationary roughness of clean
interface dynamics. The two-times density-density correlation and
mean-squared-displacement correlation age and their temporal scaling
follows a multiplicative law similar to the one found at criticality.
The linear responses also age and the comparison with their associated
correlations shows that the equilibrium fluctuation-dissipation
relation is modified in a simple manner that allows for the
identification of an effective temperature characterizing the dynamics
of the slow modes.  The effective temperature is closely related to
the vortex liquid-vortex glass crossover temperature. Interestingly
enough, our study demonstrates that the glassy dynamics in the
vortex glass is basically identical to the one of a single elastic
line in a disordered environment (with the same type of scaling though
with different parameters).  Possible extensions and the experimental
relevance of these results are also discussed.

\end{abstract}


\maketitle

\section{Introduction}\label{sec:intro}

Vortex matter in high-temperature superconductors has been extensively
studied for many years.~\cite{blat94,bran95,cohe97,natt00} In
particular, the nature of the dynamics of interacting vortices in
their different phases has attracted much attention due to its
relevance in technological applications.

Most studies of vortex systems have focused on the analysis of their
static properties.  It is well known that at low temperatures any
small amount of disorder destroys the long range order of the
Abrikosov lattice,~\cite{larkin} leading to different \textit{glassy}
phases.  For small magnetic field, {\it i.e.} small vortex density,
quasi-long range order is retained in the so-called \textit{Bragg
glass}~(BG).~\cite{grie91,cubi93,giam94,giam95} The key
characteristics of the BG are the logarithmic decay of spatial
correlations and the absence of dislocations, giving rise to a well
defined structure factor. For increasing fields, the appearance of
dislocation loops signals the transformation to a new disordered
phase, characterized by the rapid destruction of Bragg peaks observed
in neutron scattering.~\cite{cubi93} This is the so-called
\textit{vortex glass}~(VG).~\cite{fish89,fish91} Increasing the
temperature both glassy phases transform into a \textit{vortex
liquid}~(VL).  Transport and magnetic measurements showed that at
low-fields the BG solid melts into the VL through a first order
transition, while the transition from the VG solid to the VL at high
fields is continuous.~\cite{safa93,kwok94} Whether the latter
corresponds to a true phase transition or a crossover is still under
debate.~\cite{rege91,boki95,weng96,weng97,kisk98,otte98,olso00,reic00,kier00,rosenstein03,lidmar03}

Beyond the nature of the continuous VG-VL transition, important
efforts were devoted to the interpretation of the
\textit{irreversibility line}~(IRL).~\cite{mull87,yesh88,deak94} 
Experimentally, the IRL marks both the
onset of the irreversibility in zero field cooling and field cooling
magnetization measurements and the separation of two transport regimes
in the field-temperature diagram. Above the IRL, thermal energy
dominates, vortices are unable to be pinned, and any amount of current
results in a linear current-voltage characteristics. Below the IRL,
the pinning energy dominates and flux lines are irreversibly pinned.
In this case, the resistivity drops to nearly zero with a nonlinear
current-voltage characteristics. Originally, the proposition that the
IRL signals the continuous VG-VL phase transition was supported by
experiments showing the scaling of the current-voltage
characteristic,~\cite{koch89,petr00} and by simulations in randomly
frustrated 3D XY models without screening showing evidence for a
finite temperature critical point.~\cite{rege91,olso00,olsson}
However, contradictions in the experimental scaling of the
current-voltage characteristics,~\cite{jian97,strachan,petr00} on the one
hand, and the disappearance of the finite temperature transition in
the randomly frustrated 3D XY model when screening is 
restored,~\cite{boki95,weng96,weng97,kisk98} on the other hand, have raised new
questions on how to interpret the IRL.

Based on the re-examination of experimental data and the results of
numerical simulations of the over-damped dynamics of the
London-Langevin model, Reichhart {\it et al} \cite{reic00} argued that
the vortex-glass criticality is arrested at a crossover temperature.
Below this temperature, the relaxation times  as obtained from the study of
the resistivity grow very quickly in a way that is consistent with the
Vogel-Fulcher law commonly found in relaxation studies of structural
glasses.~\cite{edig00} This study suggests that the VL and 
the VG are then just separated by a crossover phenomenon in which
relaxation times go beyond the experimental time-window at a
glass temperature $T_g(H)$.
Below this crossover line one expects to find an out-of-equilibrium
 system, the VG, with all the properties of more standard
glasses.

In the last decade there has been important progress in the
understanding of the out-of-equilibrium dynamics of glasses.  A key
characteristic of relaxing glassy systems is the loss of stationarity
reflected by their \textit{aging} properties, meaning that the
system's dynamics depend on the time elapsed after the preparation of
the sample,~\cite{vincent-etal} $t_w$.  As a consequence, dynamic
correlation functions dependend now on two times, the ``waiting
time'', $t_w$, and the time $t$ elapsed during the measurement.  Also
the linear response functions of glassy systems show aging effects,
being dependent on $t_w$ and $t$, and they are anomalous in the sense
that they are not related to their associated correlation functions by
the equilibrium fluctuation-dissipation theorem (FDT).  The relation
between the two functions, correlation and response, remains, however,
rather simple.  It has been characterized in a number of glassy
systems, allowing for a kind of classification of the
out-of-equilibrium dynamics of disordered
systems.~\cite{lesuch,crisanti}  The comparison between linear
response and correlation has shed light onto the role played by
different degrees of freedom in the relaxation of the full system. It
allowed for the identification of an effective temperature in these
non-equilibrium systems with slow dynamics.~\cite{peliti}  For
example, the violation of the FDT was studied numerically in driven
vortex lattices with random pinning.~\cite{kolt02}

Metastability and aging-like phenomena have been observed in the
nonlinear transport in the VG state of single crystal
Bi$_2$Sr$_2$CaCu$_2$O$_8$ samples\cite{Sas,Portier} and analyzed in
numerical simulations.~\cite{Exartier,olson03} Also, history dependent
effects have been found in YBa$_2$Cu$_3$O$_7$ near the peak
effect.~\cite{bekeris} Aging phenomena has been reported in the
magnetization of granular Bi$_2$Sr$_2$CaCu$_2$O$_8$ samples at low
fields.~\cite{papa99,papa02} However, in this case, the effect is due
to a chiral glass phase that is found in three-dimensional Josephson
junctions networks with $\pi$ junctions,~\cite{Li01} thought to model
{\it granular} high-$T_c$ superconductors.  Theoretically, Nicodemi
and Jensen studied a simplified two-dimensional lattice Hamiltonian
model for the effective vortex dynamics to address the issue of
equilibrium and out-of-equilibrium dynamics.~\cite{nico02} They
focused on the magnetization relaxation using both one-time and 
two-times correlation functions. They observed that relaxation times
dramatically increase upon decreasing temperature, with
a Vogel-Fulcher-like dependence,
and established some analogies with glass formers and supercooled liquid
systems. This {\it two-dimensional} model emphasizes the effect
of random quenched disorder but overlooks the importance of the
three-dimensionality of flux lines and their fluctuations
along the axis parallel to the magnetic field.
Other studies focused on the out-of-equilibrium
features of vortex creep motion at low temperatures.~\cite{nico01,kolt05a}

In the present work, we compare the dynamics of vortex matter in
high-temperature superconductors with that of structural glasses
aiming at clarifying the nature of the VG phase. We extend and
complement our previous study of the out-of-equilibrium dynamics of a
London-Langevin model.~\cite{bust06} Extensive numerical simulation of
the over-damped dynamics were performed to study correlation functions
such as the roughness of the flux lines, their dynamic wandering, the
structure factor and mean-square displacement all generalized to
include two-times dependencies. Focusing on the mean-square
displacement and its conjugated response function we show that the
scaling of both functions is multiplicative and we analyze how this is
related to a non-trivial violation of the FDT.~\cite{bust06}

The outline of the paper is as follows. In Sec.~\ref{sec:back} a brief
description of the vortex phase diagram and glassy properties is
presented. Section~\ref{sec:model} gives an elaborated description of
the model and details on the numerical simulations.  The quantities of
interest for the present study are presented in
Sec.~\ref{sec:quant}. In Sec.~\ref{sec:sdk} we show how
the VG phase is approached both from the structural
and dynamical properties. Finite size effects are discussed in
Sec.~\ref{sec:finite}. The aging behavior of flux lines in the vortex glass and their
temperature dependence is presented in Sec.~\ref{sec:aging-temp}. 
The multiplicative scaling of
aging correlations and responses is described in
Sec.~\ref{sec:scaling}, the violation of the FDT, the effective
temperature and its relationship with the the dynamical arrest of the VL diffusion
are discussed. 
Finally, Sec.~\ref{sec:conc} is devoted to the discussion
of our results.

\section{Background}\label{sec:back}

In this Section we present the general background on which we base our 
study. Our aim is to give here a brief review of  the vortex matter phases in
high-temperature superconductors
as well as a description of the main features of the
non-equilibrium relaxation in glassy systems and the violation of FDT 

\subsection{Vortex phases in high-temperature superconductors}

\begin{figure}[!tbp]
\includegraphics[angle=-90,width=.48\linewidth,clip=true]{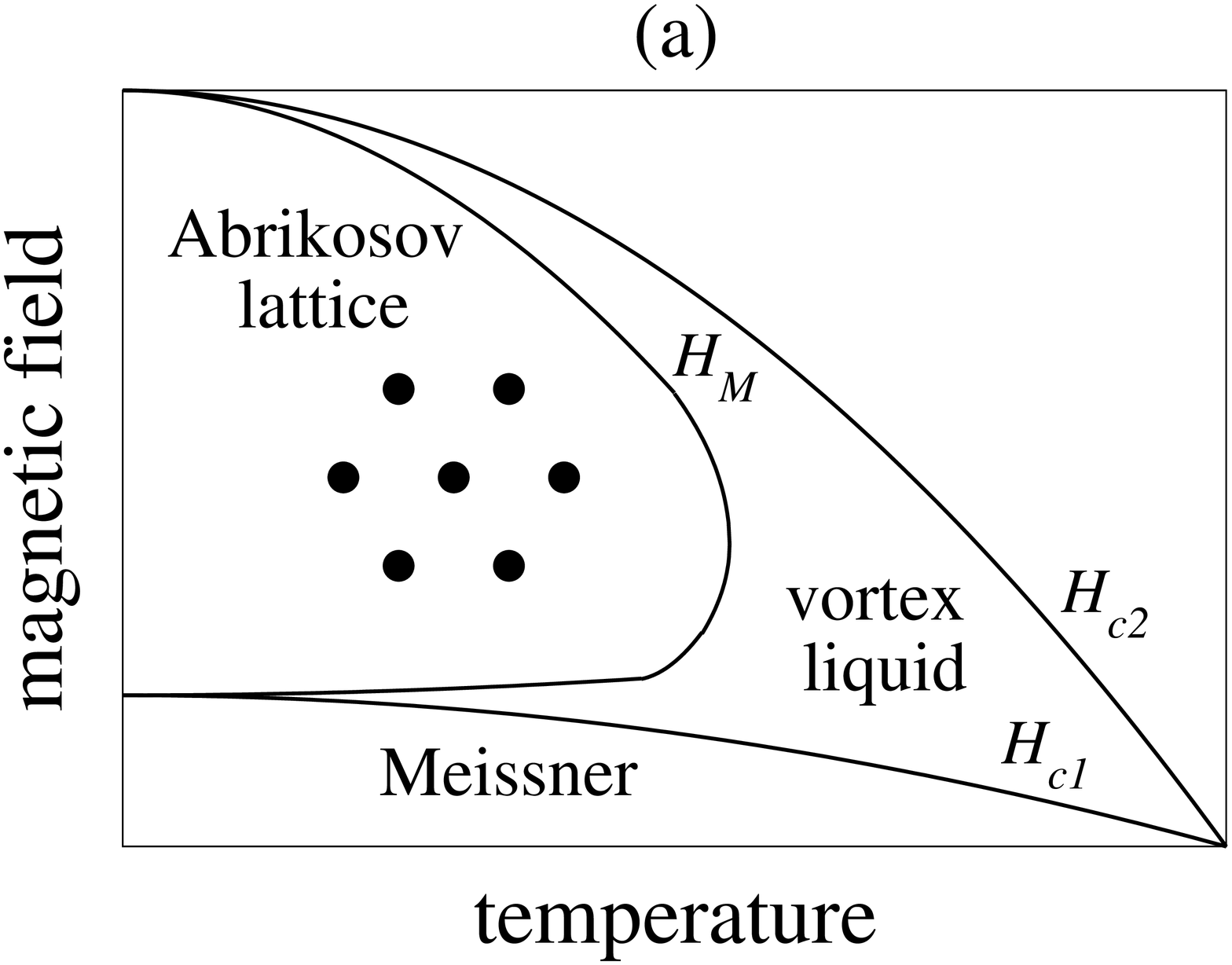}
\includegraphics[angle=-90,width=.48\linewidth,clip=true]{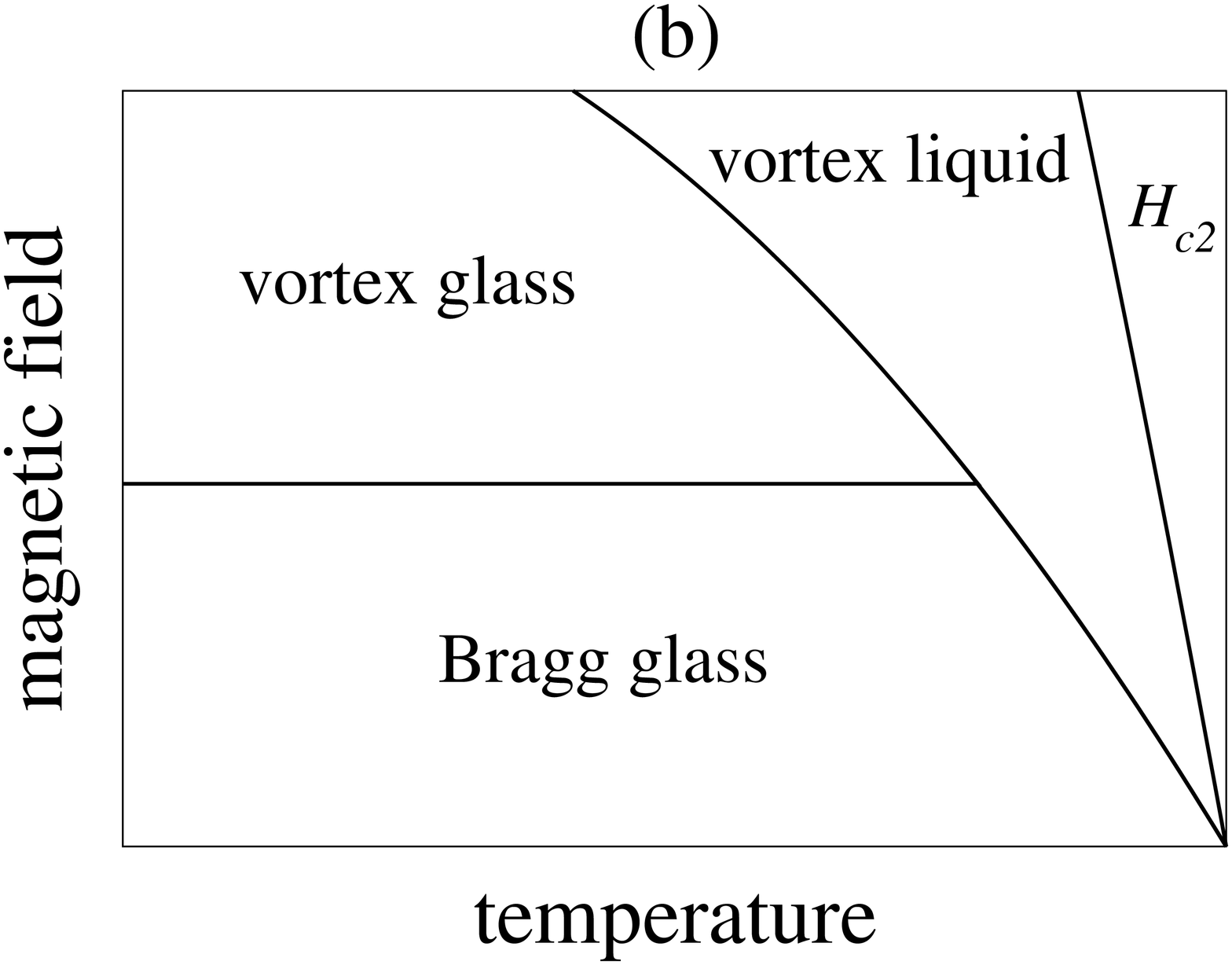}
\caption{\label{f:vgphases} Sketch of the vortex phase diagram in
high-temperature superconductors for (a) clean and (b) dirty samples. In (a)
the Abrikosov lattice phase, prototype of the vortex solid, is located between
the two critical magnetic fields, $H_{c1}$ and $H_{c2}$, at low
temperatures. At high temperatures the Abrikosov lattice melts into a vortex
liquid phase at $H_M(T)$. Below $H_{c1}$ the Meissner phase without vortex is
found. In (b) the Meissner phase at very low magnetic field is not shown and
the three phases, Bragg glass, vortex glass and vortex liquid, are sketched.}
\end{figure}

Several review articles describe the physics of vortex 
matter.~\cite{blat94,bran95,cohe97,natt00} Here we briefly summarize the main
characteristics of vortex phases in high-temperature superconductors,
referring the reader to these specialized reviews
\cite{blat94,bran95,cohe97,natt00} for further details.

In type II superconductors, 
below a lower critical field $H_{c1}$, the Meissner effect does not
allow the magnetic field to penetrate the superconductor sample. Above
$H_{c1}$, the magnetic field penetrates into the superconductor
sample, continuously increasing the flux penetration until the upper
critical field $H_{c2}$ where the sample becomes normal, see
Fig.~\ref{f:vgphases}. The magnetic field penetrates the
superconductor under the form of flux lines (or vortices) each one
carrying a flux quantum $\Phi_0=hc/2e$. Vortices are composed by an
inner normal core of radius $\xi$ (the coherent length), surrounded by
super-current screening over a distance $\lambda$ (the penetration
length).  In the absence of disorder and at low temperatures the
vortices form a triangular crystal, the so-called Abrikosov 
lattice,~\cite{abri57,degennes,tinkham} due to the repulsive vortex-vortex
interaction. At high temperature thermal fluctuations melt the Abrikosov
lattice into a vortex liquid (VL) phase when the magnetic field
$H_M < H_{c2}$ is reached. In the VL vortices move with a non-zero diffusion
constant, leading to a linear current-voltage characteristics and
dissipation. These observations have led to the phase diagram in
Fig~\ref{f:vgphases}(a).  The re-entrant behavior of the melting line
$H_M(T)$ originates in the drastic softening of the elastic constants
of the Abrikosov lattice when the vortex density is lowered close to
$H_{c1}$.

The  high-temperature superconductors reveal new vortex
physics phenomena related to the relevance of thermal fluctuations and disorder
in these systems. A schematic representation of a vortex
phase diagram, when disorder effects are included,
 in high-temperature superconductors is presented in
Fig~\ref{f:vgphases}(b).~\cite{giam97,otte98} This type of phase
diagram is experimentally observed in layered systems like
YBa$_2$Cu$_3$O$_{7-x}$ (YBCO) \cite{deli97} and
Bi$_2$Sr$_2$CaCu$_2$O$_{8+s}$.~\cite{zeld94}  In
Fig~\ref{f:vgphases}(b) the Meissner phase at very low magnetic field
is not shown. At high temperature thermal fluctuations dominate and
the VL is observed. At low temperature disorder dominates and, when it
is short-ranged correlated, two \textit{glassy phases} are observed:
the Bragg glass (BG)~\cite{Klein} at low magnetic field and the vortex
glass (VG) at high magnetic field. The BG presents a structure factor
with well defined intensity Bragg peaks at the reciprocal wave vectors of the
triangular lattice. Long-range
order, however, is destroyed by disorder leading to quasi-long range order with
spatial correlations decaying logarithmically  and power-law diverging Bragg 
peaks.~\cite{giam94,giam95} 
At high temperature the BG melts into the VL phase through a 
first-order transition analogously to
the melting of the Abrikosov lattice.
Upon increasing magnetic field or disorder, at low temperatures, the BG 
transforms into the vortex glass (VG) also through a first order
transition.~\cite{giam97,hu01,teitel01}

In the VG phase scenario, 
disorder destroys long-range order leading to a
short-range amorphous structure.~\cite{fish89,fish91} 
While the high-temperature VL can be
considered to be made of mobile vortices moving unhindered over the pinning
potential, the low temperature VG is thought to be an
immobile amorphous solid with the flux lines localized on the pinning centers
in order to minimize the pinning energy. The VG theory predicts that
collective effects are able to produce infinite energy barriers at low
temperatures, leading to a strictly zero flux-flow resistance with no
dissipation.~\cite{fish89,fish91} 
This leads to interesting scaling theories for the energies
involved in creep motion of flux lines.~\cite{feig89,natt90,kolt05a} 
Experimentally, it is very difficult to conclude
whether infinite barriers, as predicted by the VG theory,
really exist (distinguishing between infinite or very high 
barriers becomes unfeasible). Moreover, as commonly
observed in other glassy systems, large energy barriers lead to
slow relaxations and below a crossover temperature experiments
cannot be done in equilibrium.


\subsection{Aging and violation of FDT}
\label{subsec:aging}

Aging and the modification of the equilibrium FDT are two important
properties observed in experiments and simulations of glassy systems
and also captured in an analytic solutions to simple mean-field-like
models.~\cite{vincent-etal,lesuch} We here briefly describe them.

\subsubsection{Aging}

\begin{figure}[!tbp]
\includegraphics[angle=-0,width=\linewidth,clip=true]{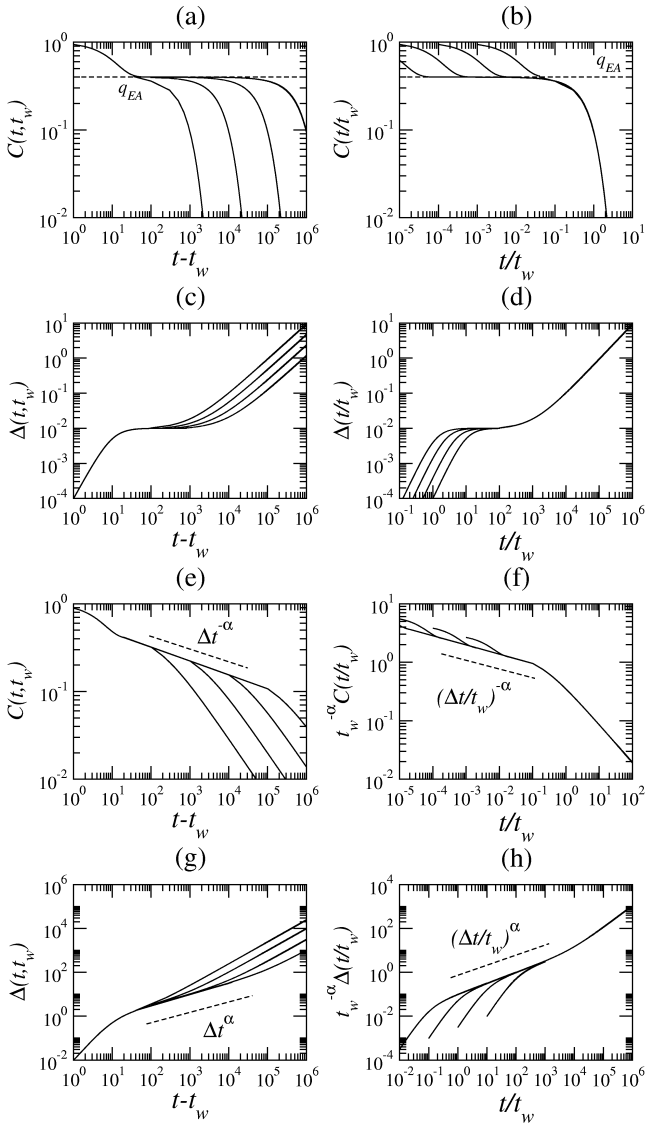}
\caption{\label{f:sketch1} Sketch of the relaxation of correlations and
displacements and their corresponding scaling forms in glassy regimes. (a)-(b)
Correlation with additive aging scaling; (c)-(d) displacement with additive
scaling; (e)-(f) correlation with multiplicative aging scaling; (g)-(h)
displacement with multiplicative scaling. Note that the stationary decays at
short time-differences in the left panels do not scale in the right
panels.}
\end{figure}

Consider the two-times global correlation function
\begin{equation}
C(t,t_w)=\frac{1}{\mathcal{N}} \langle O(t) O(t_w) \rangle ,
\label{def-corr}
\end{equation}
with $O(t)$ the global observable of interest, and $\mathcal{N}$ a
normalization factor ensuring $C(t,t)=1$. In a stationary regime the
correlation function depends only on the time difference $\Delta t
\equiv t-t_w$, {\it i.e.} $C(t,t_w)=C(\Delta t)$. Out-of-equilibrium
stationarity is not ensured. In relaxing glassy systems it is actually
lost.  The correlation then depends on both times, $t$ and $t_w$, in
such a way that the longer the waiting time the slower the
decorrelation of the system, {\it i.e.} the system is
aging. Generally, in the long $t_w$ limit, the two-times correlation
presents a separation of time scales, with an initial stationary decay
for $\Delta t \ll t_w$, depending only on $\Delta t$, followed by the aging
regime for $\Delta t \gg t_w$. This behavior is usually described by
\begin{equation}
C(t,t_w)=C_{st}(\Delta t)+C_{ag}(t,t_w),
\end{equation}
with $C_{st}(0)=1-q_{EA}$ and $C_{ag}(t,t)=q_{EA}$, and
\begin{eqnarray}
\lim_{\Delta t \rightarrow \infty} C_{st}(\Delta t) &=& 0, 
\\
\lim_{\Delta t \rightarrow \infty} \lim_{t_w \rightarrow \infty} C(t,t_w) 
&=& q_{EA}.
\end{eqnarray}
These equations define the Edwards-Anderson order parameter, $q_{EA}$. The
correlation function first decays from $1$ to $q_{EA}$ in a time-translation
invariant manner for $\Delta t \ll t_w$. At longer time differences 
the correlation decays
further from $q_{EA}$ to $0$ depending on both $t$ and $t_w$. This general
behavior of the correlation $C$ is sketched in Fig.~\ref{f:sketch1}(a).

In many cases the aging part of the correlation can be described as
$C_{ag}(t,t_w)=f\left[h(t)/h(t_w)\right]$, with $h(t)$ a system dependent
monotonic function. This form is found analytically 
in simple coarsening systems below their ordering temperature
and the $p$-spin disordered models  for fragile glasses.~\cite{lesuch} It 
also describes very accurately numerical data for Lennard-Jones 
mixtures \cite{kob00} and light-scattering 
measurements in colloidal suspensions,~\cite{viasnoff}
among other glass forming particle systems.

In Fig.~\ref{f:sketch1}(b) we present the scaling of the aging part of the
correlation shown in Fig.~\ref{f:sketch1}(a) using a scaling variable $t/t_w$:
\begin{equation}
\label{e:ag-adit}
C_{ag}(t,t_w) = \widetilde{C}_{ag}\left(\frac{t}{t_w}\right).
\end{equation}
The case $h(t) \propto t^s$ with $s$ any power 
is usually called ``simple aging''.  Note that the
short-time stationary regime does not scale in Fig.~\ref{f:sketch1}(b).
The type of behavior described by (\ref{e:ag-adit}) is known as
\textit{additive aging scaling}. 

Sometimes, the simple aging is slightly modified in favor of a
``sub-aging'' or ``super-aging'' scenario in which $h(t)$ is not just a
power law of time. In the more common former case the ratio $h(t)/h(t_w)$
can be approximated by $\Delta t/t_w^{\mu}$ with $\mu < 1$ in the short
time-difference limit. These cases are, of course, still compatible with the
additive separation in stationary and aging 
regimes  described by (\ref{e:ag-adit}).

A different, less common, functional form describing the decay of
correlations is given by the \textit{multiplicative aging scaling}:
\begin{equation}
\label{e:ag-mult}
C(t,t_w)=t_w^{-\alpha}\,\widetilde{C}_{ag}\left( \frac{t}{t_w} \right),
\end{equation}
where $\alpha > 0$ and we have assumed simple aging of the remaining
$t$ and $t_w$-dependent function. In this case the scaling function is
multiplied by a $t_w$-dependent correction that eventually makes the
aging contribution disappear. The scaling function is such that
$\widetilde{C}_{ag}(t/t_w)\sim f(\Delta t) t_w^\alpha$ for long $t_w$
in such a way that the correlation reaches, asymptotically, a
stationary regime.  The functional form (\ref{e:ag-mult}) is found
in the auto-correlation of the massless scalar field in $d>2$,~\cite{Cukupa} 
systems at criticality,~\cite{bert01,gamb05,henk05,godreche} and a
lattice model of a directed polymer in a $1+1$ dimensional random
environment.~\cite{abarrat,yosh,yosh98,bust-unp} In Figs.~\ref{f:sketch1}(e)
and \ref{f:sketch1}(f) we present a sketch of a correlation function
with multiplicative aging and its corresponding scaling form.

Note that the difference between the additive and multiplicative scaling
is that in the former the correlation function in double-logarithmic 
scale has a  well-defined plateau at a non-vanishing value separating 
stationary from aging regimes while in the later the correlation eventually 
becomes completely stationary and the aging regime disappears. 

In a  diffusive (or anomalous diffusive) 
problem, another type of correlation function is
preferred, which is simply the displacement
\begin{equation}
\Delta(t,t_w) \equiv 
\langle [O(t) - O(t_w)]^2 \rangle , 
\end{equation}
that can be rewritten in terms of the correlation 
$C(t,t_w) \equiv \langle O(t) O(t_w)\rangle$ [note that we are 
omitting here the normalization factor ${\cal N}$ 
used in (\ref{def-corr})] as
\begin{equation}
\Delta(t,t_w) = C(t,t)+C(t_w,t_w)-2 C(t,t_w).
\end{equation}
This quantity vanishes at equal times
and increases with the time difference. 
If $C(t,t)$ {\it increases} with time 
and is not bounded then the growth of 
$\Delta$ is also unbounded. For this type of correlation function it is
also possible to write down both the additive and multiplicative aging
scaling forms:
\begin{eqnarray}
\Delta(t,t_w) &=& \Delta_{st}(\Delta t) +  \widetilde{\Delta}_{ag}\left( \frac{t}{t_w} \right) 
\label{eq:add-disp}
\; , 
\\
\Delta(t,t_w) &=& t_w^\alpha \;  \widetilde{\Delta}_{ag}\left( \frac{t}{t_w} \right) 
\; , 
\label{eq:mult-disp}
\end{eqnarray}
in the simple aging case (or with a more general function h(t) 
appearing in $\Delta_{ag}$ in sub or super aging cases).
A typical displacement has initially a ballistic growth, followed by a
sub-diffusive regime proportional to $\Delta t^{\alpha}$ with $\alpha
< 1$, and a normal diffusion regime ($\alpha=0$ in the additive aging
scaling scenario). The displacement of a $D$-dimensional 
manifold relaxing in an infinite dimensional embedding space
after a quench to low-temperature has an additive separation between a 
stationary and an aging regime as in (\ref{eq:add-disp}).~\cite{Cule,Cukule} 
In the {\it one-dimensional} scalar or $N$-component field~\cite{Cukupa} and 
Sinai diffusion~\cite{Lale,Filemo,Lemofi} the displacement follows instead 
a multiplicative scaling of the form (\ref{eq:mult-disp}).
In Figs.~\ref{f:sketch1}(c)-\ref{f:sketch1}(d) and
\ref{f:sketch1}(g)-\ref{f:sketch1}(h) we present a sketch of the
displacement with additive and multiplicative scaling and their
corresponding scaling forms.

\subsubsection{The linear response}

In order to measure a linear response
the following protocol is commonly used. At $t_w$
one applies a perturbation to the system by adding the term
$\mathcal{H}_O=h O$ of intensity $h$ to the total Hamiltonian. This
perturbation is conjugated to the observable $O$. If the perturbation 
is instantaneous, the linear response is defined as
\begin{equation}
R(t,t_w) \equiv
\left. 
\frac{\delta \langle O_h(t) \rangle }{\delta h(t_w)}\right|_{h=0}.
\end{equation}
If, instead, the perturbation is held applied during the interval 
$[t_w,t]$ the integrated response
associated to the external perturbation and the observable $O$ is
\begin{equation}
\chi(t,t_w)=\frac{\langle O_h(t)\rangle - \langle O(t)\rangle }{h}
\end{equation}
where the sub-index $h$ indicates that $O$ is measured under the 
field. 
Similarly to what we explained above for the correlation and 
displacement, the time-dependence of the integrated linear response 
can be scaled in an additive or a multiplicative form:
\begin{eqnarray}
\chi(t,t_w) &=& \chi_{st}(\Delta t) + \chi_{ag}\left(\frac{t}{t_w} \right), 
\label{chi-add}
\\
\chi(t,t_w) &=& t_w^{\alpha} \;
\widetilde{\chi}_{ag}\left(\frac{t}{t_w} \right), 
\label{chi-mult}
\end{eqnarray}
respectively, where we assumed simple aging.

\subsubsection{The fluctuation-dissipation theorem (FDT)}

In equilibrium the integrated response is stationary, $\chi(t,t_w)=\chi(\Delta
t)$, and the fluctuation-dissipation theorem (FDT) implies
\begin{eqnarray}
\frac{d C(\Delta t)}{d \Delta t} &=& - k_B T R(\Delta t), 
\label{eq:fdt1}
\\
C(\Delta t)-C(0)   &=& - k_B T \chi(\Delta t),
\label{eq:fdt2}
\end{eqnarray}
($\Delta t\geq 0$)
relating correlation and instantaneous or integrated 
responses only through the temperature of the environment.  
In equilibrium the equal-times correlation is
constant, and the displacement is simply related to the correlation:
$\Delta(\Delta t) = 2[1-C(\Delta t)]$. Equations~(\ref{eq:fdt1}) and 
(\ref{eq:fdt2}) then imply 
\begin{eqnarray}
\frac{d \Delta(\Delta t)}{d \Delta t} &=& 2k_B T R(\Delta t), 
\label{eq:fdt3}
\\
\Delta(\Delta t) &=& 2k_B T \chi(\Delta t). 
\label{eq:fdt4}
\end{eqnarray}

In an aging 
out-of-equilibrium regime the integrated response also depends on two times,
see Eqs.~(\ref{chi-add}) and (\ref{chi-mult}).
A simple generalization of the FDT (\ref{eq:fdt1}) reads
\begin{equation}
\theta(t-t_w)
\frac{\partial }{\partial t_w} C(t,t_w)=k_B T_{\rm eff}(t,t_w) R(t,t_w),
\end{equation}
which gives a definition of a two-times dependent effective temperature 
$T_{\rm eff}$. It has been shown that this definition respects the
 expected properties
of a temperature for systems with slow dynamics and a bounded 
energy.~\cite{peliti} Note that the integration of the above relation over 
a time-interval $[t_w,t]$ can take a very complicated form if $T_{\rm eff}$
depends on times in an involved manner. If, as it turns out to
be in systems with additive scaling, 
it is piecewise constant one finds  a linear 
relation between the displacement and the integrated linear response
in each interval. Systems with multiplicative scaling also allow for the 
identification of a simple $T_{\rm eff}$ once the factors $t_w^\alpha$ 
are taken into account. 

\begin{figure}[!tbp]
\includegraphics[angle=-90,width=\linewidth,clip=true]{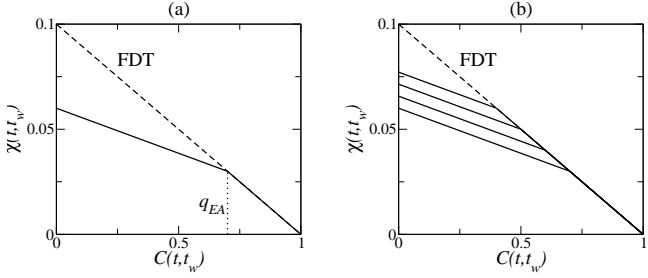}
\caption{\label{f:sketchFDT} Sketch of the fluctuation-dissipation 
relation between integrated response and correlation: (a) additive scaling 
with a stable breaking point at $q_{EA}$; (b) multiplicative scaling
with a breaking point that drifts towards $(0,1/T)$ in the limit 
$t_w\to\infty$.}
\end{figure}

The parametric plots $\chi(C)$ and $\chi(\Delta)$ have become a useful
tool to analyze the violation of FDT.~\cite{crisanti,lesuch} For
systems in equilibrium these plots are just straight lines with slope
$-1/k_BT$ and $1/2k_BT$, respectively. For glassy systems evolving out
of equilibrium one has to distinguish those with an additive from
those with a multiplicative scaling.  For the former, the parametric
plot $\chi(C)$ [$\chi(\Delta)$] shows a $-1/k_BT$ [$1/2k_B T$] slope
in the quasi-equilibrium short time-difference regime, while at long
time-differences -- for correlation values below the plateau at
$q_{EA}$ -- the slope changes to $-1/k_BT^*$ [$1/2k_BT^*$].  In the
asymptotic, $t_w\to\infty$ limit the breaking point in $\chi(C)$ occurs at a fixed
point $[q_{EA}, \chi_{EA}=1/(k_BT)(1-q_{EA})]$. This case is
schematically represented in Fig.~\ref{f:sketchFDT}(a).  In the case
of a system with multiplicative scaling one needs to eliminate the
factors $t_w^{-\alpha}$ in the decaying correlation ($t_w^\alpha$ in
the diffusing displacement) and $t_w^\alpha$ in the integrated
response in order to get a stable parametric plot. In other words,
tracing $\chi(C)$ [$\chi(\Delta)$] one gets a similar broken line with
slopes $-1/k_BT$ and $-1/k_BT^*$ [$1/2k_BT$ and $1/2k_BT^*$] but with
a breaking point that moves towards $(0,1/k_B T)$ [$(\infty,\infty)$] with
increasing $t_w$.  This behaviour is schematically represented in
Fig.~\ref{f:sketchFDT}(b). Constructing instead $t_w^{-\alpha}
\chi(t_w^\alpha C)$ or, equivalently, $t_w^{-\alpha}
\chi(t_w^{-\alpha} \Delta)$ a stable breaking point is obtained and a
parametric plot similar to Fig.~\ref{f:sketchFDT}(a) is recovered.

\section{Model}\label{sec:model}

We model vortices in superconductors as a set of elastic lines with an
interaction potential $K_0(r/\lambda)$ screened at the scale of the London
penetration depth $\lambda$. This model was introduced in the theory of flux
lattice melting by D. Nelson,~\cite{nels88,nels89} it was used to develop the
theory of the Bose glass phase,~\cite{nels92} and it has been used by several
authors in the past and until present.~\cite{ryu96a,otte98,nordborg98,lopa04,kier05}
There is now a general consensus in that this model is correct for moderately
anisotropic superconductors like YBCO. Numerical simulations find good
quantitative agreement with experimental results for this 
system.~\cite{ryu96a,otte98} We have chosen to work with such a well-known 
model to build upon previously acquired knowledge and addressing 
questions that have not been considered in the literature yet.

\subsection{Model Hamiltonian}

We consider a model for 3D elastic flux lines in a high-temperature
superconductor. The model is composed by $L$ planes labeled with index $z$ and
placed a distance $d_z$ apart. The direction of the magnetic field $B$ is 
perpendicular to the planes. The magnetic field
fixes the vortex density $n_B=B/\Phi_0$, where $\Phi_0=hc/2e$ is the flux
quantum perpendicular to the planes. The system contains $N$ flux lines,
where the $i$-th flux line is characterized by coordinates ${\bf
r}_i(z)$, which means that each flux line is composed by $L$ elements with
two-dimensional in-plane coordinates ${\bf r}_i(z)=[x_i(z),y_i(z)]$. These elements are
sometimes called ``vortex pancakes''.~\cite{ryu96a} The superconductor
has anisotropy $\epsilon=\xi_c/\xi_{ab}= \lambda_{ab}/\lambda_c$, with
$\xi_{ab}$ and $\xi_c$ the coherence lengths, and $\lambda_{ab}$ and
$\lambda_c$ the penetration depths. The axes are such that $c\parallel z$ and
$ab\parallel {\bf r}_i$.

Assuming that ${\bf r}_i(z)$ varies slowly with $z$
 the Hamiltonian for a London model of elastic flux lines is ${\cal H}= \sum_{z}
{\cal H}_{z}$, with
\begin{equation}
\label{e:hamil}
{\cal H}_{z} = \sum_i
U_{l}(\Delta{\bf r}_{iz}) + U_{d}({\bf r}_{iz}) + \sum_{i<j}U_{in}({\bf r}_{jz}-{\bf
r}_{iz}),
\end{equation}
where $U_l$, $U_d$, and $U_{in}$ are the elastic line energy between
two pancakes of the same flux line, the interaction of a flux line
with a quenched disorder potential, and the in-plane interaction
energy between two different flux lines, respectively.  Let us now
describe each term in some detail.

The repulsive interaction energy between line elements belonging to different
flux lines and in the same plane $z$ is
approximated as \cite{reic00,otte98,ryu96a}
\begin{equation}
\label{e:K0}
U_{in}= 2 \epsilon_0 d_z K_0(r/\lambda_{ab}),
\end{equation}
where $K_0$ is the modified Bessel function, and with
\begin{equation}
\epsilon_0=(\Phi_0/4\pi\lambda_{ab})^2.
\end{equation}

The elastic line energy of the $i$-th vortex  is
\begin{equation}
U_{l}= \frac{1}{2}c_{l} (\partial_z {\bf r}_i)^2 dz,
\end{equation}
where $c_l$ measures the line tension. A natural choice for $d_z$ is the
distance between CuO planes, in which case one can account for the effect of
the Josephson coupling \cite{ryu96a} using
\begin{equation}U_{l}=c_{l}\lambda_J
|\partial_z {\bf r}_i| \text{\,\,\, for \,\,\,} |\Delta{\bf
r}|
> 2 \lambda_J
\end{equation}
while keeping the previous expression for
$| \Delta{\bf r}| < 2 \lambda_J$, where $\lambda_J = d_z/\epsilon$
is the Josephson length, and $\Delta{\bf r}_{iz}={\bf r}_{i,z+1}-{\bf r}_{i,z}$
measures the separation between two adjacent flux line elements.

The quenched disorder potential due to the impurities is \cite{blat94}
\begin{equation}
\label{e:Udes}
U_{d}({\bf r})=\int d^2{\bf r'} \; u({\bf r}') p(|{\bf r}-{\bf
r}'|),
\end{equation}
where the form factor is
\begin{equation}
p(r)= 2\xi_{ab}^2/(r^2+2 \xi_{ab}^2),
\end{equation}
and
\begin{equation}
\label{e:udescorr}
\langle u({\bf r},z) u({\bf r}',z') \rangle = \gamma\delta({\bf
r}-{\bf r}')\delta_{zz'}
\end{equation}
defines the disorder strength~\cite{otte98,blat94} $\gamma$.


We model the dynamics of the vortex system with the Langevin equation
\begin{equation}
\label{e:langevin}
\eta\frac{\partial {\bf r}_{iz}(t)}{\partial t} = -\frac{\delta
{\cal H}[\{ {\bf r}_{iz}(t)  \}]}{\delta {\bf r}_{iz}}+{\bf
f}^T_{iz}(t) \; ,
\end{equation}
where  $\eta$ is the Bardeen-Stephen friction coefficient.
The thermal force ${\bf f}^T_{iz}(t)$
satisfies
\begin{equation}
\langle f^T_{iz,\mu}(t) \rangle = 0,
\end{equation}
and
\begin{equation}
\langle f^T_{iz,\mu}(t) f^T_{i'z',\mu'}(t') \rangle = 2\eta k_BT
\delta(t-t') \delta_{zz'} \delta_{ii'}\delta_{\mu
\mu'},
\end{equation}
where $\mu,\mu'=x,y$ and $T$ is the thermal
bath temperature.

\subsection{Numerical details}

There is no standard numerical procedure
to compute the in-plane interaction (\ref{e:K0}) with periodic boundary
conditions. Ryu and Stroud worked with a summation calculation
over image vortices,~\cite{ryu96a} while Zim\'anyi and 
coworkers~\cite{otte98,otte00,reic00} 
and Nordborg and Blatter \cite{nordborg98} used instead
the periodic extension of a simplified Fourier representation. 
Here we describe the numerical procedure we
used to compute the terms contributing to the Hamiltonian (\ref{e:hamil}).

\begin{figure}[!tbp]
\includegraphics[angle=0,width=\linewidth,clip=true]{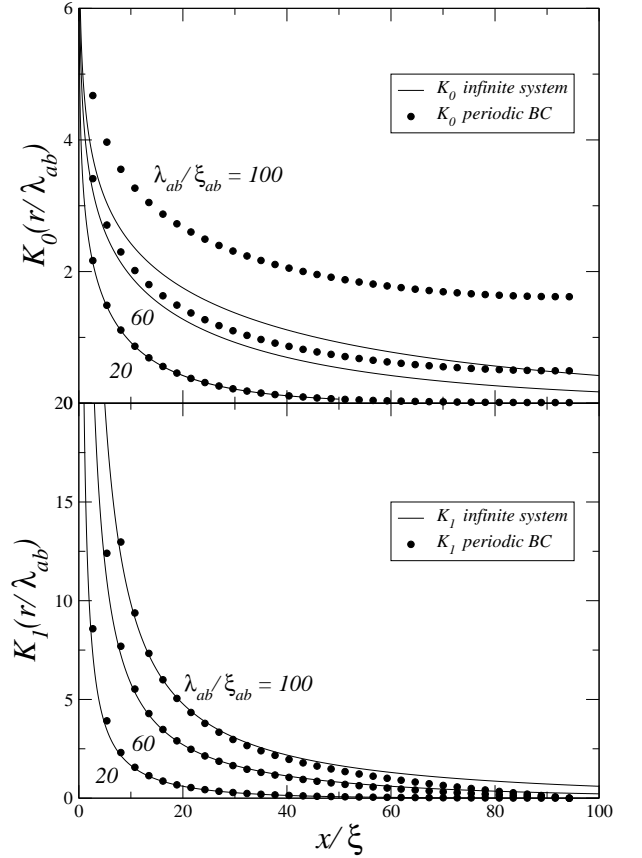}
\caption{\label{f:K0K1Bessel} Comparison between the analytic Bessel functions
$K_0(x)$ and $K_1(x)$ and the numerical results using periodic boundary
conditions and $L_x/\xi_{ab} = 188$. Data is shown up to $x=L_x/2$.}
\end{figure}

The $z$-planes have dimensions $L_x \times L_y$ and we used periodic boundary
conditions to minimize finite size effects in the planes. To compute the
in-plane interaction each direction in the $z$-planes is discretized in
$N_{x(y)}$ points a distance $a_{x(y)}$ apart, such that $L_{x(y)}=N_{x(y)}
a_{x(y)}$ for the $x(y)$-direction. 
The $K_0$ dependence of the interaction potential, Eqs.(19), corresponds
to the case of vortices in infinite samples. In a finite sample with
periodic boundary conditions, the effective potential that takes into
account the effect of the summation over periodic images, $\tilde{K}_0(r)$,
has to be calculated.
Instead of calculating $\tilde{K}_0$ in real
space, we found more convenient to work in the reciprocal space, using a
discrete Fourier transform of the in-plane interaction (\ref{e:K0})
that accounts for the periodic boundary condition. This reads
\begin{equation}
\label{e:K1}
\tilde{U}_{in} = 
\frac{2 \epsilon_0 d_z \Phi_0}{L_x L_y}\,\,
\frac{1}{\lambda_{ab}^2 h^2_{k,l}-1},
\end{equation}
with $k$ and $l$ the indices for the discrete reciprocal space, and
\begin{equation}
h^2_{kl}=\frac{2}{a_x^2}
\left[\cos\left(\frac{2 \pi k}{N_x}\right)-1\right]+
\frac{2}{a_y^2}\left[\cos\left(\frac{2 \pi l}{N_y}\right)-1\right].
\end{equation}
The above expressions account explicitly for the periodic boundary
conditions in $L_x$ and $L_y$.
Using 
The numerical value of $\tilde{U}_{in}$ is easily computed in a $kl$
discrete mesh. To calculate the real space value of $\tilde{U}_{in}$, a fast Fourier
transform FFTW~\cite{FFTW} routine is used. In Fig.~\ref{f:K0K1Bessel} we show
the $K_0$ and $K_1$ Bessel functions corresponding to the infinite system,
and the effective functions $\tilde{K}_0$ and $\tilde{K}_1$ 
that consider the periodic boundary conditions, obtained with this method.
 Different values of $\lambda_{ab}$ are shown to follow the general trend. As can be
observed, the new functions take into account the periodic boundary
conditions: the  $\tilde{K}_1$ function, which is proportional to the force
between vortex pancakes, goes to zero at the edge of the simulation box
($L_x/2$), which is a necessary condition when working with periodic boundary
conditions. 
Finally the in-plane force, proportional to $\tilde{K}_1(r/\lambda_{ab})$,
derived from this potential energy is stored in a force table during the
computational run.

Since the $z$-direction is discretized in $L$ planes, which naturally
correspond to the CuO planes in a model superconductor, the elastic
interaction is discretized as
\begin{equation}
U_{l} (\Delta{\bf r}_{iz})= \left\{
\begin{array}{ll}
\frac{1}{2} c_{l} \left(\frac{\Delta {\bf r}_{iz}}{d_z}\right)^2 d_z,
& \qquad \text{for \,\,\,} |\Delta{\bf r}_{iz}| > 2 \lambda_J, \\
\\
c_{l}\lambda_J \frac{|\Delta {\bf r}_{iz}|}{d_z},
& \qquad \text{for \,\,\,} |\Delta{\bf r}_{iz}| > 2 \lambda_J.
\end{array}
\right.
\end{equation}

\begin{figure}[!tbp]
\includegraphics[angle=-90,width=\linewidth,clip=true]{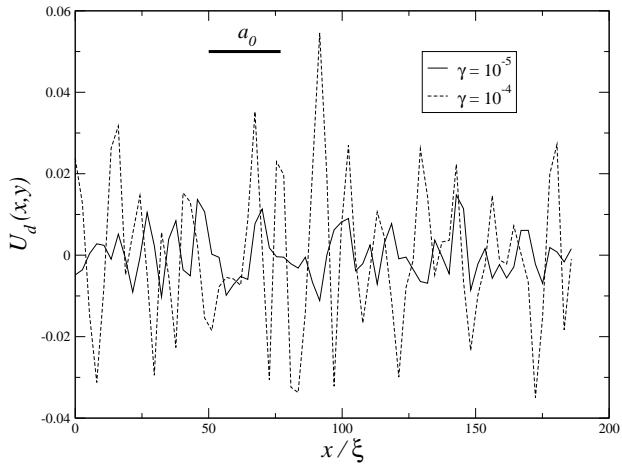}
\caption{\label{f:udes} Quenched disorder potential for two values of the 
strength $\gamma$. The Abrikosov length $a_0 = 26.935$ is shown for
reference.}
\end{figure}

In order to compute the disorder potential (\ref{e:Udes}) we use the same
discrete mesh as for the in-plane interaction. The discrete disorder potential
at a given mesh point, indexed by $mn$, is given by
\begin{equation}
U_d=2 \sqrt{\frac{24 \sqrt{2}\pi^2 a_x a_y\gamma}{d_z}}
\sum_{m'=1}^{N_x} \sum_{n'=1}^{N_y} u'_{m'n'}\,p_{m'n'},
\end{equation}
where $u'_{mn}$ is a random number uniformly distributed between $0$ and $1$
and thus the prefactors enforce Eq. (\ref{e:udescorr}) to hold. The discrete
form factor is given by
\begin{equation}
p_{mn}=\sum_{m'=1}^{N_x}\sum_{n'=1}^{N_y}
\left[(m-m')^2 a_x^2+(n-n')^2 a_y^2+2\right]^{-1}.
\end{equation}
The indices $m$ and $n$ run over the discrete mesh, {\it i.e.} $m=1,2,\ldots,N_x$
and $n=1,2,\ldots,N_y$. In Fig.~\ref{f:udes} a typical disorder potential
realization is
shown for two intensities, $\gamma = 10^{-4}$ and $10^{-5}$, later
used in the simulations. The force exerted by the disorder potential at a
given point of the mesh is stored in a force table.

Since the in-plane and disorder potentials are evaluated in a discrete mesh
but the pancake positions are continuous variables in the Langevin dynamics
equations, a four-point interpolation method is used.

\subsection{Parameters of the simulation}

The above model gives a good quantitative description of the vortex
physics of moderately anisotropic high-temperature superconductors
like YBCO.~\cite{otte98,ryu96a,blat94,bscco-clem,bscco-olson} We
therefore choose the values of the parameters to mimic this system:
$\epsilon=1/5$, $\lambda_{ab}/\xi_{ab}=100$, $\lambda_J/\xi_{ab}=16$,
and $c_{l}=\epsilon^2\epsilon_02\left[1+\ln
(\lambda_{ab}/d_z)\right]/\pi$.~\cite{ryu96a} The strength of disorder
is set to $\gamma = 10^{-5}$, for which case we find that above
$B_{cr}\sim 0.002 H_{c2}$ the system is in the VG at low temperatures
(see Section \ref{sec:sdk}). Time is normalized by $t_0=\xi_{ab}^2
\eta/\epsilon_0$, length by the vortex lattice parameter (or Abrikosov
length) $a_0=[2\Phi_0/(\sqrt{3}B)]^{1/2}$, energy by $\epsilon_0d_z$,
and temperature by $\epsilon_0d_z/k_B$. We simulate $N=56$ vortices in
a box of size $7a_0\times 4\sqrt{3}a_0$ with periodic boundary
conditions for the in-plane coordinates. The $z$ direction is
discretized in $L=50$ planes with free boundary conditions. Averages
are performed over $10$ realizations of disorder.

\section{Quantities of interest}\label{sec:quant}


In order to study the out-of-equilibrium dynamics we use a
two-times protocol. First the system is equilibrated at an initial high
temperature well inside the VL ($T_i=0.3$) evolving during $t=10^4$
steps. Then the system is quenched to a low temperature $T$, where the
time count is set to zero. Starting with this far from equilibrium
initial condition, the system is let to evolve during a waiting time
$t_w$, after which the quantities of interest are measured.

With the aim of fully characterizing the dynamical properties of the vortex 
system we study several $t$ and $t_w$ dependent quantities.
First we focus on the study of finite size effects by monitoring 
the roughness of the lines, a quantity that is 
used in the study of interfaces dynamics,~\cite{barabasi}
here generalized to include the $t$ and $t_w$ dependence:
\begin{equation}
\label{e:w2}
\langle w^2(t,t_w) \rangle = 
\frac{1}{LN} \sum_{iz} 
\left \langle \left[ \delta x_{iz}(t) - 
\delta x_{iz}(t_w) \right] ^2 \right \rangle,
\end{equation}
where $\delta x_{iz}(t)=x_{iz}(t)-\overline{x_{i}(t)}$ accounts for the
displacement of the $iz$ line segment relative to the center of mass of the
$i$-th elastic line, $\overline{x_{i}(t)}=L^{-1}\sum_z x_{iz}(t)$, along one 
of the two axis of the two-dimensional planes on which the pancakes move. Another
quantity that is 
useful to analyze finite size effects is the dynamic
wandering,~\cite{yosh-unp} $W(z,t,t_w)$, defined as
\begin{equation}
\label{e:wandering}
W(z,t,t_w)=\frac{1}{N}
\sum_i \left \langle \left[\Delta \mathbf{r}_{iz}(t)-
\Delta \mathbf{r}_{iz}(t_w)\right]^2 \right\rangle,
\end{equation}
with $\Delta \mathbf{r}_{iz}(t)=\mathbf{r}_{iz}(t)-\mathbf{r}_{i0}(t)$. This
quantity measures how the displacement of the vortex pancakes (or line
elements) in the $z$-th plane correlates with the bottom plane $z=0$ between
$t$ and $t_w$. Although this was defined for the general vortex model, it is
worth mentioning that when in-plane interactions and disorder potential are
absent, {\it i.e.} for the Edwards-Wilkinson (EW) equation,~\cite{EW82} 
the dynamic wandering
is related to the height-height correlation function studied in surface growth
phenomena.~\cite{barabasi,yang92,he92,yang93}

Aging in these systems is best characterized by monitoring the 
 mean-square-displacement (MSD) of the pancakes in the planes
\begin{equation}
B(t,t_w)=\frac{1}{LN}\sum_{iz} 
\left\langle [x_{iz}(t)-x_{iz}(t_w)]^2 \right\rangle,
\label{def:MSD}
\end{equation}
and the two-times wave-vector dependent 
density-density correlation function
\begin{equation}
\label{e:corrC}
C_k(t,t_w)=\frac{1}{LN}\sum_{iz} 
\left\langle e^{-ik\,[x_{iz}(t)-x_{iz}(t_w)]} \right\rangle
\end{equation}
that has been analysed in great detail in relaxing
glass forming liquids.~\cite{kob00} The mean-square displacement 
$B$ is of the form of the generic displacements $\Delta$ discussed 
and the density-density correlation is of the form of the 
generic decaying correlation $C$ discussed in Sec.~\ref{subsec:aging}.

In order to complete the characterization of the aging properties and 
to study the modifications of the FDT in the out-of-equilibrium
regime, we measure a linear response function by applying a random force of
the form ${\bf f}_{iz}= \delta s_{iz} \hat{{\bf x}}$ at time 
$t_w$ on a replica of the system, where $\delta$ is the intensity of the
perturbation, and $s_{iz}=\pm1$ with equal probability.~\cite{parisi,kolt02} 
The integrated response is
\begin{equation}
\chi(t,t_w)= \frac{1}{LN\delta}
\sum_{lz} \left\langle s_{iz}[x^{\delta}_{iz}(t)-x_{iz}(t)] \right\rangle,
\end{equation}
where $x^{\delta}_{iz}$ and $x_{iz}$ correspond to the position evaluated in
two replicas of the system, with and without the perturbation. 
The equilibrium FDT implies
\begin{equation}
B(\Delta t)=2k_BT \chi(\Delta t),
\end{equation}
where the $\Delta t=t-t_w$ argument represents stationarity. In the aging regime the
FDT is violated and one constructs the modified relation
\begin{equation}
\label{e:BXfdt}
B(t,t_w)=2k_BT_{\rm eff}(t,t_w) \chi(t,t_w).
\end{equation}

\section{Vortex glass}\label{sec:sdk}

The vortex phase diagram of a model very similar to the one studied in
this work was obtained by Zim\'anyi and co-workers,~\cite{otte98} and
it has the qualitative behavior shown in Fig.~\ref{f:sketch1}(b). In
order to show that we are in the VG region of the phase diagram for
the present model superconductor, we study the behavior of the
in-plane vortex structure factor and the behavior of
dynamic correlations and relaxation times in the VL. 

\subsection{Structure factor: loss of crystalline order}

\begin{figure}[!tbp]
\includegraphics[angle=-90,width=\linewidth,clip=true]{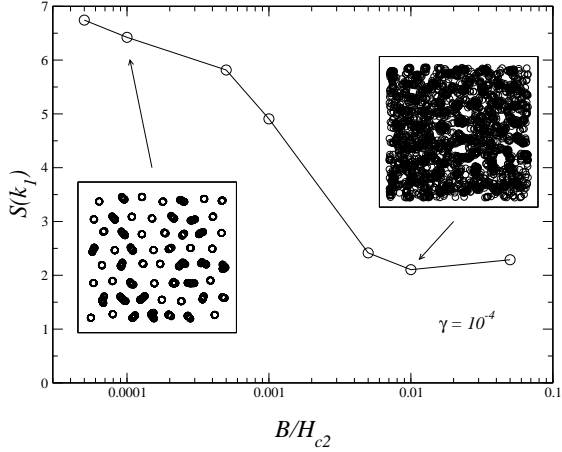}
\caption{\label{f:sdk} Structure factor at $k_1$, the main peak of the
Abrikosov lattice, and for the disorder intensity $\gamma = 10^{-5}$
as a function of the vortex density parametrized by the field
$B$. Insets: a number of superposed snapshots of the system.}
\end{figure}

In Fig.~\ref{f:sdk} the vortex
structure factor at $k_1$, the principal peak of the Abrikosov lattice
with parameter $a_0$, is shown while varying the magnetic field
intensity $B$ (proportional to the vortex density) and at fixed
disorder intensity $\gamma = 10^{-5}$. For this case we find that
above $B_{cr}\sim 0.002 H_{c2}$ the Bragg peaks disappear and the flux
lines are frozen in a highly amorphous structure at low
temperatures. Furthermore, $B_{cr}$ is dependent on the
disorder intensity $\gamma$. We therefore choose to study the case
with $B=0.01 H_{c2} \gg B_{cr}$, which is deep within the VG regime at
low $T$ for both $\gamma = 10^{-4}$ and $10^{-5}$. The region with
higher structure factor $S(k_1)$ below $B_{cr}$ presumably corresponds to a BG phase,
but to be sure one should test the finite size behavior of the spatial
correlations and structure factor. 
The insets clearly show the structural differences
between two snapshots with different densities. These snapshots are
projections of the two-dimensional pancake coordinates of the $L$
planes into the $z=0$ plane.

\subsection{Diffusion in the VL and dynamical arrest}\label{sec:diff}

Once working at fields $B > B_{cr}$,
upon increasing temperature the system 
should go through  the VG-VL crossover  line. 
This could be numerically observed as
a sudden increase of the vortex diffusion. However, since the ``pancake'' diffusion
is sub-linear \cite{ryu96a} in the VL the data analysis and the identification of 
the crossover line may be subtle.  
This can be better analyzed starting at high temperatures, from the
VL side, and studying the relaxation of dynamical correlation functions

\begin{figure}[!tbp]
\includegraphics[angle=-90,width=\linewidth,clip=true]{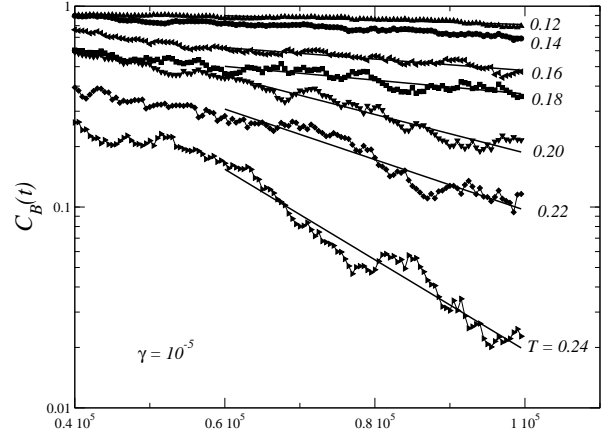}
\caption{\label{f:CdB2} Correlation $C_B(t)$ for different temperatures 
and $\gamma =10^{-5}$. Each curve is labeled by the
temperature. The corresponding exponential fits are shown.}
\end{figure}

\begin{figure}[!tbp]
\includegraphics[angle=-90,width=\linewidth,clip=true]{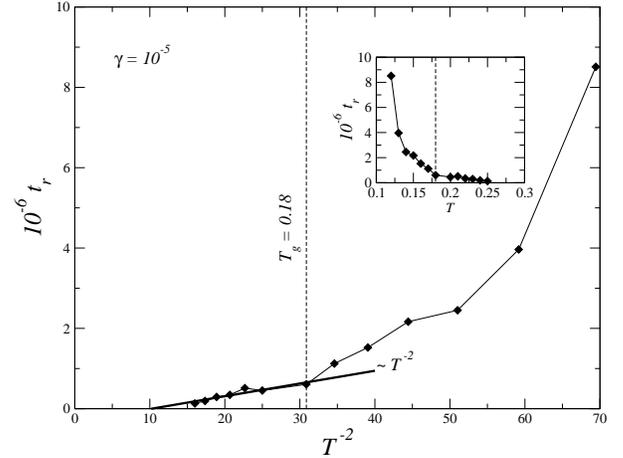}
\caption{\label{f:t0dT} Relaxation times $t_r$ obtained from the exponential
fit $C_B(t)=A e^{-t/t_r}$. The crossover temperature $T_g$, where the VL
relaxation time departs from the single line behavior $t_r \propto 1/T^2$ is
quoted. The inset shows the sudden increase in $t_r$ with decreasing
temperature.}
\end{figure}

In order to clarify the nature of the dynamic arrest and to define a 
crossover glass temperature $T_g$,
we therefore study the change in diffusion upon decreasing the temperature in the VL
phase, by analyzing the MSD defined in Eq.~(\ref{def:MSD}).

At high temperatures, well inside the VL, the system can  be easily equilibrated and the MSD
depends on $\Delta t = t - t_w$, $B(t,t_w)\equiv B(\Delta t)$.
Two regimes of pancake diffusion can be observed in this case. For
times longer than a saturation time, $\Delta t > t_x$, diffusion is
normal, $B(\Delta t) \sim \Delta t$, corresponding to the diffusion
of the center of mass of each line. 
At shorter times, $\Delta t < t_x$, the diffusion of the pancakes
corresponds to that of line segments in a line in which the longitudinal
correlation length is still growing. This is the same as considering an
infinite line $L \rightarrow \infty$ and it is characterized by
sub-diffusion,~\cite{ryu96a} $B(\Delta t) \sim \Delta t^{1/2}$.

One aims to extract a characteristic relaxation time from the diffusion
evolution. To this end, Zim\'anyi and coworkers \cite{otte98} used an
exponential fit, $C(t) = A e^{-t/t_r}$, to describe the correlation
\begin{equation}
C(t)=\exp \{ -\overline{\langle [ \mathbf{r}(t)-\langle \mathbf{r}(t)\rangle ]^2\rangle } \},
\end{equation}
where $\mathbf{r}(t)$ measures the two-dimensional in-plane displacements of the 
flux line elements.
However, this is only consistent with a normal diffusion regime, and should be
carefully interpreted. In their simulations they used an equilibration time
of $10^5$ and measured $C(t)$ for the following $10^5$ time interval. This is
the same as using a waiting time of $10^5$ from the given initial
condition. The exponential fit is valid if $t_x < 10^5$, corresponding to
center of mass diffusion.

As mentioned above, Ryu and Stroud \cite{ryu96a} studied the melting
of the VL using a different model and observed that the pancake
diffusion is sub-linear, $B(\Delta t) \sim \Delta t^{\alpha}$ with the
exponent $\alpha=1/2$ at high temperatures and $\alpha < 1/2$ at low
temperatures.  This is consistent with our results.

Within the same scheme, we calculate the evolution of the correlation for the
largest waiting time used in our simulations, $t_w=10^5$,
\begin{equation}
\label{e:CdB2}
C_B(t)=\exp \{ - [B(t,t_w=10^5)]^2\},
\end{equation}
and fit it to the form $C_B(t)=A e^{-t/t_r}$, that follows after using
the sub-linear exponent $\alpha = 1/2$. In Fig.~\ref{f:CdB2} we show
this correlation in a log-linear representation for different
temperatures and $\gamma = 10^{-5}$. The fits for $\Delta t >
0.6\times 10^5$ are also shown. From this data we obtain the
relaxation time $t_r$ presented in Fig.~\ref{f:t0dT}. The inset shows
that there is a sudden increase in the relaxation time upon decreasing
temperature. For non-interacting flux lines the temperature dependence
of the relaxation time should be $t_r \propto
1/T^2$,~\cite{blat94,ryu96a} and the same is expected deep in the VL.
The main panel in Fig.~\ref{f:t0dT} shows that indeed for high
temperatures $t_r$ grows linearly with $1/T^2$. Below a temperature
$T_g \approx 0.18$ we observe that the relaxation time $t_r$ departs
strongly from the $1/T^2$-dependence and presents a remarkable
growth. Furthermore, for $T<T_g$ is not possible to equilibrate the
system and correlation functions depend on the two times, $t$ and
$t_w$, as we will analyze in Sec.~\ref{sec:aging-temp}. Therefore, the
temperature $T_g$ signals the crossover from the equilibrium
high-temperature regime to the out-of-equilibrium low-temperature
regime.

\section{Finite size effects}\label{sec:finite}

Before analyzing the out-of-equilibrium behavior of different
correlation and response functions of the VG, it is necessary to
previously identify finite size effects in the dynamics of the system.
We here compare finite size effects in the single elastic line
dynamics with those in the much more complex systems we are dealing
with, a set of interacting lines in the presence of quenched
disorder. To this end we study both the two-times roughness
\eqref{e:w2} and the dynamic wandering \eqref{e:wandering}. We also
show how finite size effects appear in two other quantities of
interest, the pancake MSD \eqref{def:MSD} and the density-density
correlation function \eqref{e:corrC}.

\subsection{Single lines: the Edwards-Wilkinson equation}

\begin{figure}[!tbp]
\includegraphics[angle=0,width=\linewidth,clip=true]{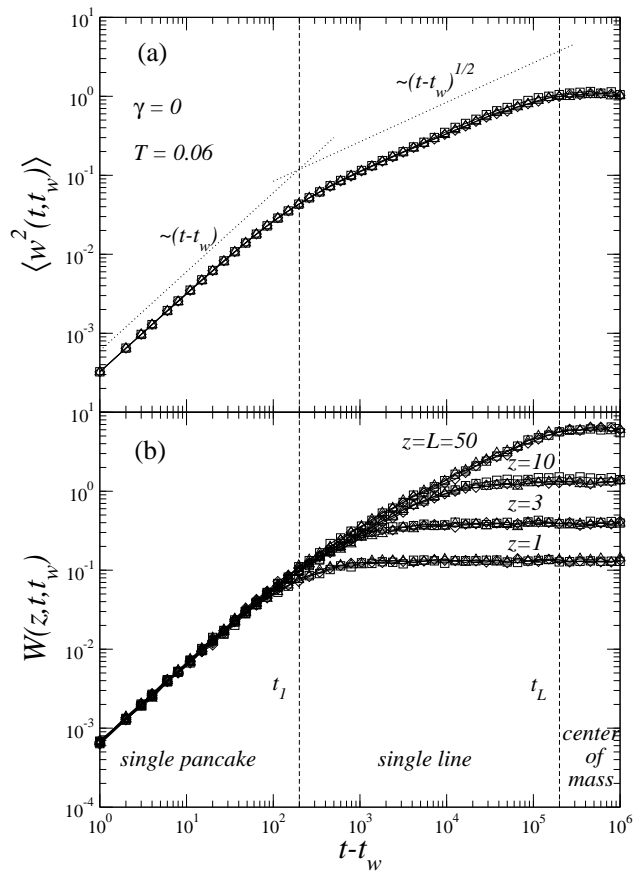}
\caption{\label{f:w2t-Wzt_425} (a) Roughness and (b) dynamic wandering 
of non-interacting elastic lines without quenched disorder 
(EW elastic lines) at $T=0.06$. Three dynamic regimes are
highlighted, corresponding to single pancake, single line, and center of mass
dynamics. Note that different curves with $t_w = 10^4,\,10^5$ and $10^6$ (open
squares, open diamonds and open triangles, respectively) overlap, showing that
these quantities are in a stationary regime.}
\end{figure}

\begin{figure}[!tbp]
\includegraphics[angle=-90,width=\linewidth,clip=true]{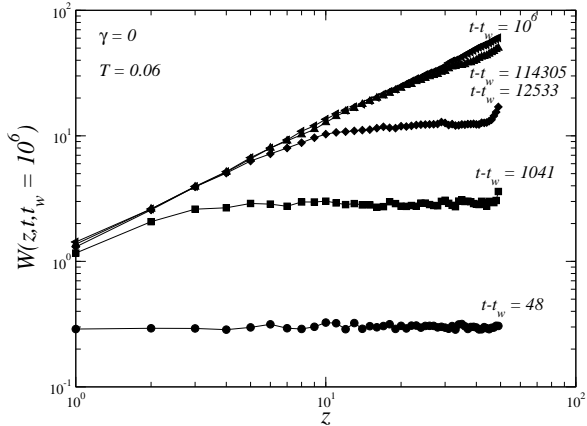}
\caption{\label{f:Wzt-z_425} Dynamic wandering as a function of the plane
index $z$. $T=0.06$, $L=50$ each curve is labeled by $\Delta t$ and the
waiting time is $t_w =10^6$ for all. Since $W(z,\Delta t)$ saturates at a value
proportional to the longitudinal correlation length $\xi_{||}$, the growth of $\xi_{||}$ is
observed for increasing $\Delta t$ at fixed $z$.}
\end{figure}

\begin{figure}[!tbp]
\includegraphics[angle=0,width=\linewidth,clip=true]{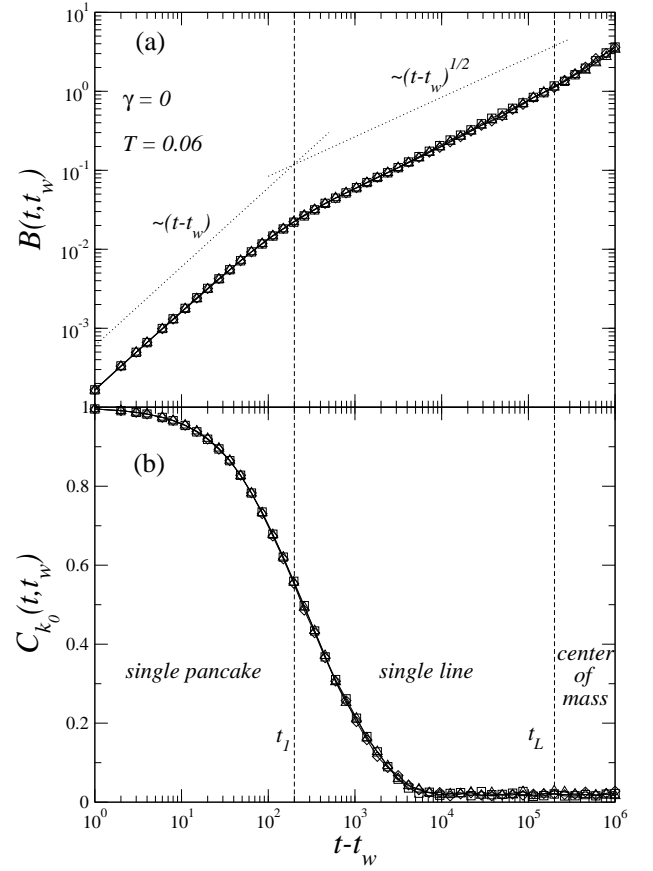}
\caption{\label{f:Bdt-Cdt_425} (a) Pancake MSD $B$ and (b) correlation
$C_{k_0}$ for the same parameters as in Figs.~\ref{f:w2t-Wzt_425} and
\ref{f:Wzt-z_425}. $k_0$ is the wave vector at the first peak of the structure
factor.  The three dynamic regimes are clear in $B$, but the correlation
$C_{k_0}$ does not present any signature of them.  Symbols are as in
Fig.~\ref{f:w2t-Wzt_425}.}
\end{figure}

When the in-plane interactions and disorder potential are absent from
the Hamiltonian (\ref{e:hamil}), Eq.~(\ref{e:langevin}) reduces to $N$
decoupled Edwards-Wilkinson (EW) equations.~\cite{EW82} The EW
equation was studied in great detail in the context of growing
interface dynamics.~\cite{barabasi} After a transient the roughness of
the EW interface becomes stationary and it
follows the Family-Vicsek scaling \cite{fami85}
\begin{equation}
\label{e:Family-Vicsek}
\left \langle w^2(\Delta t)  \right\rangle = 
L^{2 \zeta} f\left( \frac{\Delta t}{t_x} \right),
\end{equation}
where $t_x \sim L^\theta$ is the saturation time, $L$ is the system size, and
$f(x)$ is a scaling function with $f(x) \sim x^{2\beta}$ for $x \ll 1$ and
$f(x) \sim cons.$ for $x \gg 1$. The growth and roughness exponents, $\beta$
and $\zeta$ respectively, are related to the dynamic exponent $\theta$ through
the scaling relation $\theta=\zeta/\beta$. The exponents take values
$\zeta = 1/2$, $\beta = 1/4$, and $\theta = 2$. This scaling
form clearly states how the system size is involved in the line dynamics.
Other quantities, such as the global correlation function, can 
however be non-stationary and show aging.~\cite{Cukupa} 

In order to check how the size and time dependence of our elastic lines
compare to the EW ones, we set disorder and in-plane interactions to zero and
we analyse the full $\Delta t$ dependence of the roughness and dynamic
wandering once the stationary regime has been reached.  In
Fig.~\ref{f:w2t-Wzt_425}(a) we show the time-evolution of the roughness. Three
regimes are clear in the figure.  For $\Delta t < t_1$ the roughness grows as
$\langle w^2 \rangle \sim \Delta t$ which corresponds to a regime without
elastic interactions between the line segments. For $\Delta t > t_L$ the
roughness saturates to an $L$-dependent value, while for $t_1 < \Delta t <
t_L$ the roughness grows as $\langle w^2 \rangle \sim \Delta t ^{1/2}$. The
last two regimes are the ones in the scaling function of the Family-Vicsek
scaling (\ref{e:Family-Vicsek}), where $t_x \equiv t_L$.

To better understand the meaning of a characteristic distance-dependent 
time $t_z$ let us analyze the
behavior of the dynamic wandering in Fig.~\ref{f:w2t-Wzt_425}(b). For
$\Delta t < t_1$ the dynamic wandering is independent of $z$ and grows
as $\sim \Delta t$. This is the regime in which the line segment in one
plane evolves independently of the line segment in any other plane,
{\it i.e.} there is no truly elastic interaction yet. For $\Delta t >
t_1$ the line segments belonging to two adjacent planes are
elastically interacting and their relative position saturates to a
given value. However, at this time, line segments separated by more
than a single plane are still uncorrelated. This corresponds to the
growth of a longitudinal correlation length, $\xi_{||}$, along the
$z$-direction. This correlation grows as $\xi_{||} (\Delta t)\sim
\Delta t^{1/\theta}$ for $\Delta t \ll t_x$ and saturates at the
system size,~\cite{barabasi} $\xi_{||} (\Delta t)\sim L$, for $\Delta
t \gg t_x$.  The time $t_z$ signals the moment when the correlation
length is of the order of $z$, {\it i.e.} $\xi_{||}(t_z)\sim z$. In
Fig.~\ref{f:w2t-Wzt_425}(b) the behavior of $W(z,t,t_w)$ is presented
for $z=1,3,10$ and $z=L=50$, where it is shown that the dynamic
wandering is essentially a measure of the longitudinal correlation
length (the short time $\Delta t^{1/2}$ regime is not completely developed
since the system size is not big enough).

In Fig.~\ref{f:Wzt-z_425} we show the $z$-dependence of the dynamic
wandering. Since there is no $t_w$-dependence, we show here different curves
labeled by the corresponding $\Delta t$ value for a single waiting time
$t_w=10^6$. For $\Delta t < t_1$, the dynamic wandering does not depend on $z$
since line segments are independent of each other. For $t_1 < \Delta t < t_x$
two regimes are observed. For $z<\xi_{||}(\Delta t)$ one has $W \sim z$, while
for $z>\xi_{||}(\Delta t)$ the dynamic wandering saturates at a $\Delta
t$-dependent value. The final growth near $z=L$ is due to the free boundary
conditions of the line. Finally, for $\Delta t > t_x$ all the line is
correlated and $W \sim z$ for all $z$.

The growth of the longitudinal correlation length is also clear in other
quantities. For instance, in Fig.~\ref{f:Bdt-Cdt_425}(a) we show the pancake
MSD $B(t,t_w)$, where the three dynamic regimes are also observed. For $\Delta
t < t_1$ independent pancake diffusion is observed and diffusion is normal,
$B(\Delta t) \sim \Delta t$. For $t_1 < \Delta t < t_x$ diffusion is
characterized by a sub-linear law $B(\Delta t) \sim \Delta t^{1/2}$,
corresponding to the single line regime where every pancake elastically feels
the others.~\cite{ryu96a,blat94} Finally, for $\Delta t > t_x$,
when the correlation length is of the order of the system size, a trend to
normal diffusion corresponding to the center of mass diffusion is observed.

Lastly, in Fig.~\ref{f:Bdt-Cdt_425}(b) the density-density correlation
function $C_{k_0}(t,t_w)$ is shown, where $k_0$ is the position of the first
maximum of the static structure factor. This correlation shows how the
position of the pancakes decorrelate in time. Note that around $\Delta t \sim
10^4$ the pancakes have lost all information about the position at 
$t_w$ ($\Delta t=0$) and $C_{k_0}$ has effectively decayed to zero. 
There is no signature of the three dynamic regimes in the decay of 
the correlation function. 

In this subsection we demonstrated how the system size is involved in
different dynamical quantities. We also identified three main dynamic
regimes. In the following subsections we study how these features 
are reflected in the VL and VG.

\subsection{The vortex liquid}

\begin{figure}[!tbp]
\includegraphics[angle=0,width=\linewidth,clip=true]{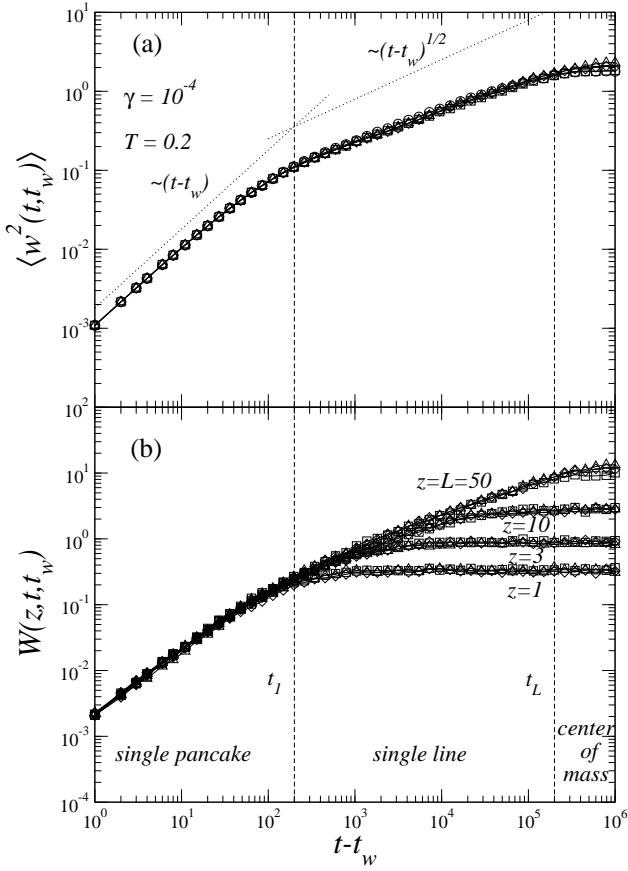}
\caption{\label{f:w2t-Wzt_306} 
(a) Roughness and (b) dynamic wandering 
in the VL 
at $T=0.2$ with disorder intensity $\gamma = 10^{-4}$
(cfr. Fig.~\ref{f:w2t-Wzt_425} where the same measurements for the 
EW line are shown).}
\end{figure}

Here we return to the case with in-plane interactions and disorder,
and present results at high temperature, $T=0.2$, and with disorder
intensity $\gamma = 10^{-4}$, well inside the VL. The analysis of
finite size effects in the roughness and dynamic wandering
(Fig.~\ref{f:w2t-Wzt_306}) shows that these are similar to the ones in
the EW line.  The values of $t_z$ are of the same order. At this high
temperature and for this disorder intensity one can argue that thermal
fluctuations are greater than disorder induced fluctuations, and the
disorder potential in (\ref{e:hamil}) can be basically
neglected. Moreover, although in-plane interactions are present, at
the working vortex density they do not much affect the dynamics
leading to essentially the same finite size effects as for the EW
line. At higher densities in-plane interactions should become the most
relevant interaction of the problem. (As an example, a change in the
scaling exponent was found with hard-core line interactions.~\cite{peta04})

\subsection{The vortex glass}

\begin{figure}[!tbp]
\includegraphics[angle=0,width=\linewidth,clip=true]{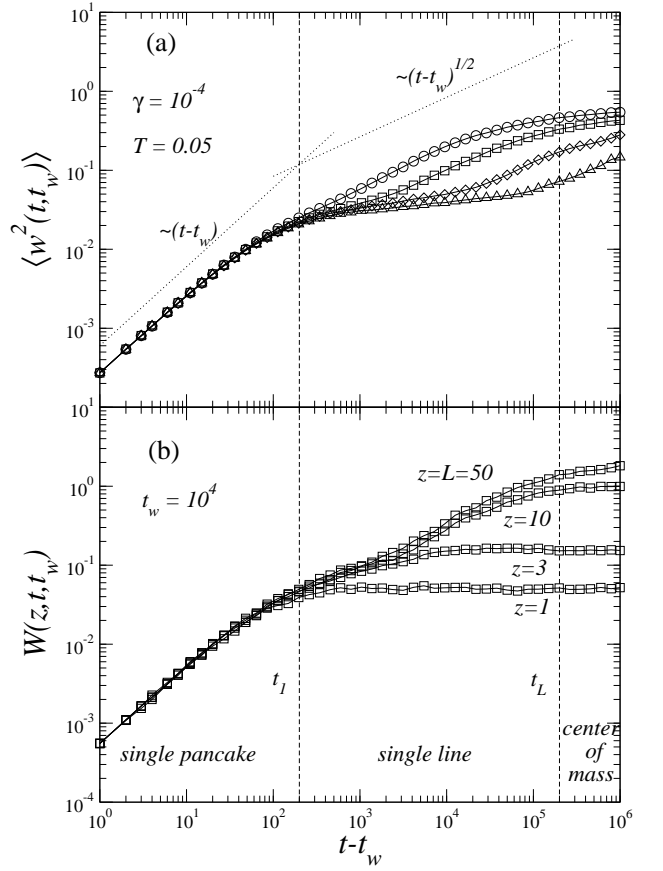}
\caption{\label{f:w2t-Wzt_325} (a) Roughness and (b) dynamic wandering in the
VG at $T=0.05$ with disorder intensity $\gamma = 10^{-4}$
(cfr. Figs.~\ref{f:w2t-Wzt_425} and~\ref{f:w2t-Wzt_306}) where the same
measurements in the EW line and the VL, respectively, are shown).  In (b) the
dynamic wandering is shown only for $t_w=10^4$.}
\end{figure}

\begin{figure}[!tbp]
\includegraphics[angle=-90,width=\linewidth,clip=true]{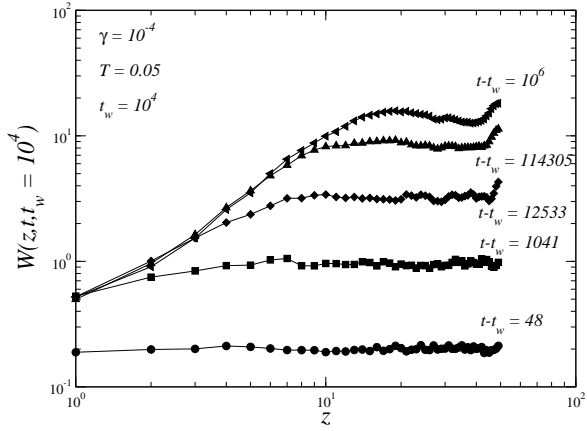}
\caption{\label{f:Wzt-z_325_tw1} Dynamic wandering as a function of the plane
index $z$. At this low temperature and for this disorder 
strength the system is aging. We show data for $t_w =10^4$
and each curve is labeled by $\Delta t$. A change in the slope of $W$ at short
$z$ is apparent. For this value of $t_w$ the longitudinal correlation length
$\xi_{||}(\Delta t=10^6)$ does not reach the system size $L$.}
\end{figure}

Here we analyze how the finite size effects reflect in the low temperature
phase. We show data for $T=0.05$ and $\gamma = 10^{-4}$. We anticipate
that aging effects appear at low temperature, but we postpone their discussion
to Sec.~\ref{sec:aging-temp}. In
Fig.~\ref{f:w2t-Wzt_325} the roughness and dynamic wandering evolution are
shown. The characteristic aging $t_w$-dependence is evident. The values of
$t_1$ and $t_L$ used in this figure are the same as in the EW case,
Fig.~\ref{f:w2t-Wzt_425}. The first stationary single pancake regime is still
present. Aging develops in the single line regime but saturation seems to take
place at the same value $t_L$, suggesting that the saturation time $t_x$ does
not strongly depend on temperature and disorder intensity for the studied
parameter values. The dynamic wandering also presents aging and the same
saturation time is apparent. In Fig.~\ref{f:Wzt-z_325_tw1} the $z$-dependence
of the dynamic wandering for $t_w=10^4$ is shown. Note that for this
value of $t_w$ and the maximum $\Delta t=10^6$ reached, the longitudinal
correlation length does not reach the system size $L$, and then the $\Delta
t=10^6$ curve shows a saturation regime for high $z$. Although the same trend
as in the EW case is present, a difference is observed in the
$\xi_{||}(\Delta t) < z$ regime. For instance, the curves with $\Delta
t=114305$ and $\Delta t=10^6$ show a change in the slope. This is
related to the fact that the roughness exponent $\zeta$ of the Family-Vicsek
scaling (\ref{e:Family-Vicsek}) changes with disorder.~\cite{barabasi,healy} In
the presence of disorder, at a given temperature there exists a characteristic
length $z_c(T)$ separating two values of $\zeta$. For $z < z_c(T)$ one has
$\zeta = \zeta_T$, while for $z > z_c(T)$ one has $\zeta = \zeta_D$, with
$\zeta_T < \zeta_D$, where $\zeta_T$ and $\zeta_D$ are the thermal (EW) and
disorder scaling exponents, respectively.

From the above considerations we choose to work in the time regime
corresponding to both $\Delta t$ and $t_w$ being shorter than
$t_x$. This ensures that aging features do not overlap saturation
effects and finite size effects should be irrelevant. For $L=50$ this
corresponds to times $\Delta t,t_w \le 10^4$.

\section{Aging in the vortex glass}\label{sec:aging-temp}

In this Section we describe aging features observed in the present
simulations. As shown in Sec.~\ref{sec:diff}, when entering
the VG phase, for $T < T_g$, is not
possible to equilibrate the system, and correlation and response
functions depend on the two times, $t$ and $t_w$, showing aging.
We start by describing the $t_w$ and $k$-dependence of the
correlation $C_k(t,t_w)$. Then we discuss the aging characteristics
of the pancake MSD $B(t,t_w)$ and its associated integrated linear response
$\chi(t,t_w)$.

\subsection{$C_k(t,t_w)$}

\begin{figure}[!tbp]
\includegraphics[angle=-90,width=\linewidth,clip=true]{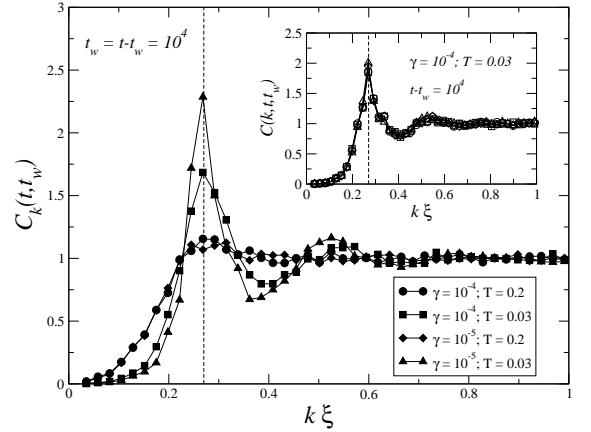}
\caption{\label{f:Cdk_301-368-418-401} Normalized static structure factor
$C_k(t,t_w)$ for $\Delta t = t_w = 10^4$ and different $T$ and $\gamma$ given
in the key. The value $k_0=0.27$ corresponds to the first peak. The inset
shows $C_{k}(t,t_w)$ for $\Delta t =10^4$ and different waiting-times, 
$t_w=10,\,10^2,\,10^3$ and $10^4$, with $\gamma = 10^{-4}$ and $T=0.03$. The
structure factor does not vary significantly with waiting time while the
system is aging.}
\end{figure}

\begin{figure}[!tbp]
\includegraphics[angle=-90,width=\linewidth,clip=true]{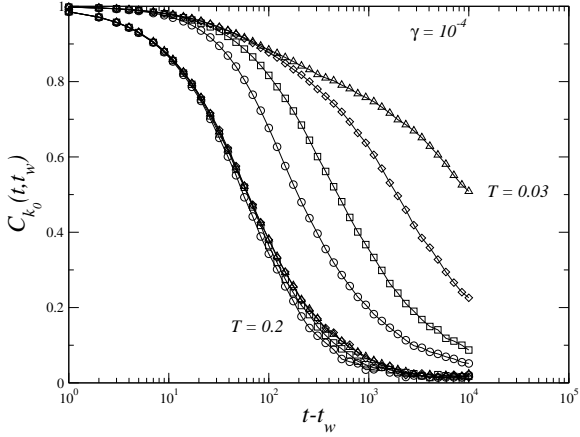}
\caption{\label{f:Cdt_301-368} Correlation $C_{k_0}(t,t_w)$ for $\gamma =
10^{-4}$ and two temperatures, $T=0.2$ and $T=0.03$. Different waiting times
are shown: circles, squares, diamonds and triangles correspond to
$t_w=10,\,10^2,\,10^3$ and $10^4$, respectively.}
\end{figure}

\begin{figure}[!tbp]
\includegraphics[angle=-90,width=\linewidth,clip=true]{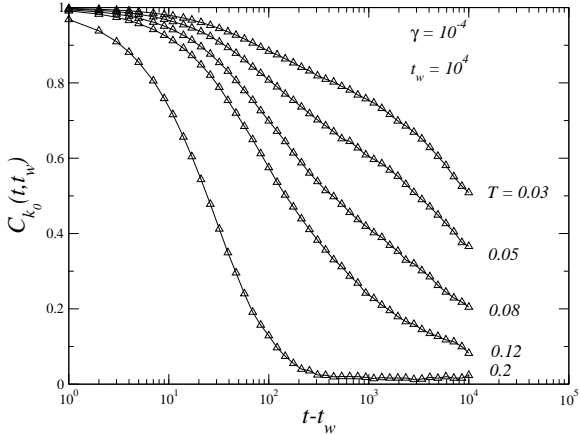}
\caption{\label{f:Cdt_tw4_301-371-312-324-368} Correlation $C_{k_0}(t,t_w)$
for fixed $t_w=10^4$ and $\gamma = 10^{-4}$. Different curves correspond to 
different temperatures given in the figure.}
\end{figure}

\begin{figure}[!tbp]
\includegraphics[angle=-90,width=\linewidth,clip=true]{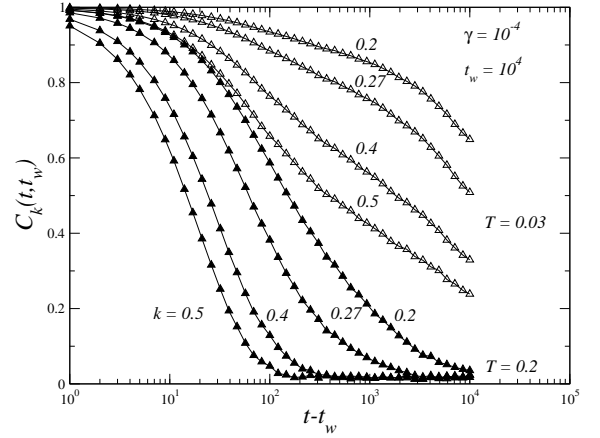}
\caption{\label{f:Cdt_k_301-368} Correlation $C_k(t,t_w)$ for fixed $t_w=10^4$
and two temperatures, $T=0.2$ and $T=0.03$ with $\gamma = 10^{-4}$. Curves for
different wave-vector values $k$ are shown.}
\end{figure}

Let us first present the $k$-dependence of the normalized static structure
factor $C_k(t,t_w)$ for fixed $\Delta t$ and $t_w$. In
Fig.~\ref{f:Cdk_301-368-418-401} we present data for $\Delta t = t_w = 10^4$
and different $T$ and $\gamma$ given in the key. At low temperature, in the VG
phase, the curves present a well defined first maximum, followed by an
observable second peak, but clearly not showing the presence of an ordered
structure. At high temperature, well in the VL phase, the first peak is not as
sharp as in the VG case, but it is clearly observed. For reference, the first
peak corresponds to $k_0=0.27$. This is the value that we used in
Fig.~\ref{f:Bdt-Cdt_425}. In the inset of Fig.~\ref{f:Cdk_301-368-418-401},
the evolution of $C_k(t,t_w)$ for fixed $\Delta t$ with waiting time is
shown. Data correspond to $\gamma = 10^{-4}$ and $T=0.03$. 
The structure factor is not strongly modified while the system is aging,
{\it i.e.} while the relaxation time increases for increasing $t_w$ (see
Fig.~\ref{f:Cdt_301-368} below). This feature is also observed in simulations
of structural glass formers using Lennard-Jones potentials.~\cite{kob00}

Figure \ref{f:Cdt_301-368} shows the evolution of the correlation
$C_{k_0}(t,t_w)$ at two temperatures, $T=0.2 > T_g$ and $T=0.03 < T_g$, and for
different waiting times. The disorder intensity is $\gamma =
10^{-4}$. At high temperature the system decorrelates rapidly and the
curves become stationary, {\it i.e.} they do not depend on $t_w$. At
low temperature a short time-difference quasi-equilibrium regime is
present, but a clear $t_w$-dependence is developed at longer
time-differences.  The system decorrelates slower at longer waiting
times, {\it i.e.} the system ages. For the longer $t_w$ value the
curve does not clearly present a plateau as it usually does in other
glassy systems.~\cite{crisanti} An additive scaling scenario would
require that the plateau develops at a longer $t_w$. To exclude this
possibility we should reach a longer $t_w$ while still being in the
$t_w < t_x$ regime. This demands to use larger system sizes $L$. We
believe, however, that this is not necessary since 
the correlation actually decays according to a
multiplicative scaling.  We give further support to this proposal
below.

The temperature dependence of $C_{k_0}(t,t_w)$ for $t_w=10^4$ is shown in
Fig.~\ref{f:Cdt_tw4_301-371-312-324-368}.  At the highest temperature the
system is in equilibrium but for the remaining values it evolves out
of equilibrium. The relaxation time decreases
for decreasing temperature (at fixed $t_w$). 

Figure \ref{f:Cdt_k_301-368} shows the $k$-dependence of $C_k(t,t_w)$ for
$t_w=10^4$ and two temperatures, $T=0.2$ and $T=0.03$. At both temperatures
the system decorrelates faster with decreasing wave-vector $k$, meaning that
the relaxation is much faster on small than on large length scales. This is
exactly the same behavior observed in the Lennard-Jones glass-forming liquid
out of equilibrium by Kob and Barrat.~\cite{kob00}

\subsection{$B(t,t_w)$}

\begin{figure}[!tbp]
\includegraphics[angle=0,width=\linewidth,clip=true]{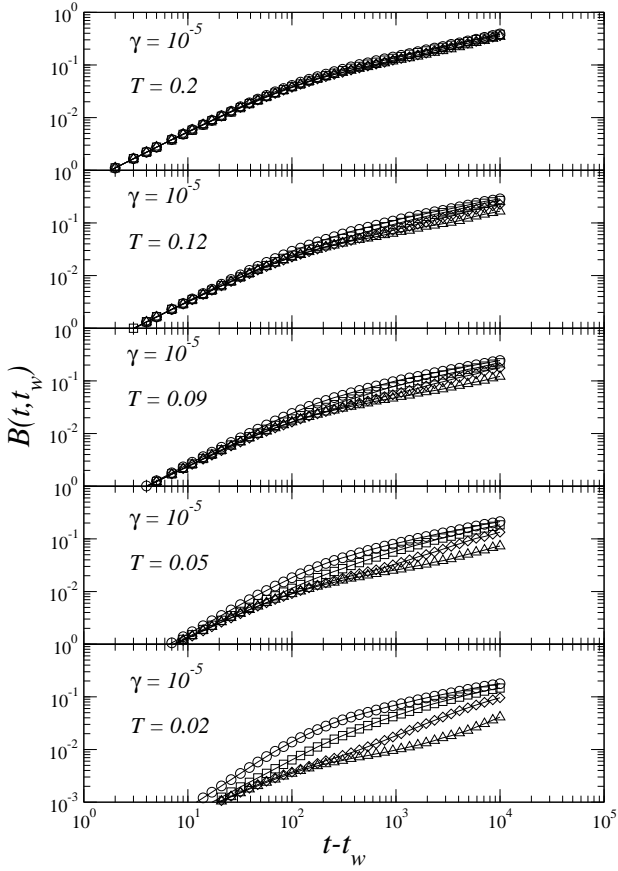}
\caption{\label{f:Bdt_b_418-410-407-403-400} Pancake MSD $B(t,t_w)$ at
different temperatures. Circles, squares, diamonds and triangles
correspond to $t_w=10,\,10^2,\,10^3$ and $10^4$, respectively.}
\end{figure}

\begin{figure}[!tbp]
\includegraphics[angle=0,width=\linewidth,clip=true]{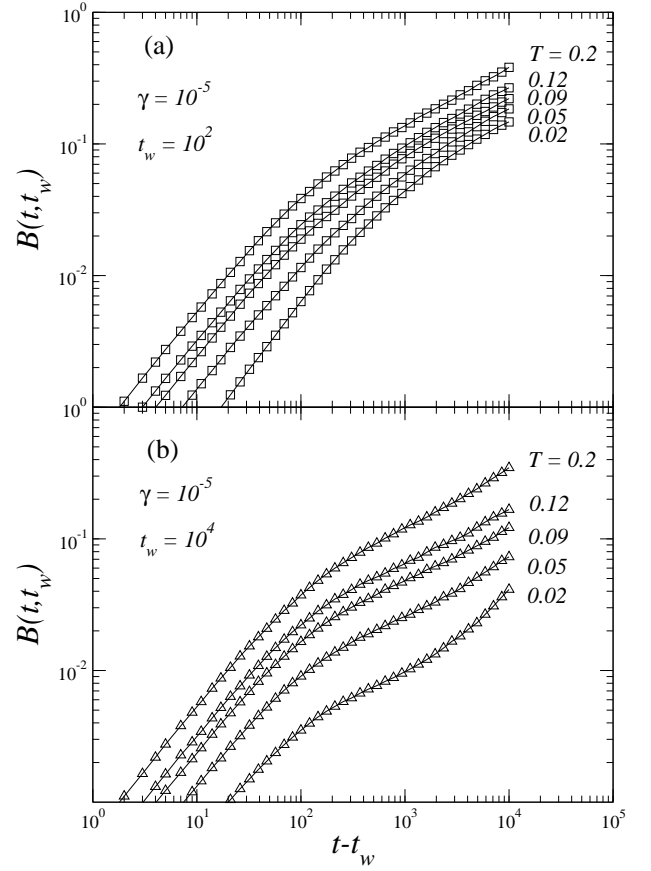}
\caption{\label{f:Bdt_418-410-407-403-400} $B(t,t_w)$ for different
temperatures and for two waiting times: (a) $t_w=10^2$ and (b) $t_w=10^4$.}
\end{figure}

In Fig.~\ref{f:Bdt_b_418-410-407-403-400} the time and temperature evolution
of the pancake MSD \cite{bust06} $B(t,t_w)$ is shown for different waiting
times and fixed disorder intensity $\gamma = 10^{-5}$. $B(t,t_w)$ is
stationary for the high temperature value $T=0.2$. For temperatures lower
than $T_g \approx 0.18$ the
aging behavior develops. As temperature is decreased the separation between
curves with different waiting times increases, indicating that the system is
deeper in the aging regime. Figure \ref{f:Bdt_418-410-407-403-400} shows the
temperature dependence of $B(t,t_w)$ for two waiting times, $t_w = 10^2$ in
Fig.~\ref{f:Bdt_418-410-407-403-400}(a) and $t_w = 10^4$ in
Fig.~\ref{f:Bdt_418-410-407-403-400}(b). After the initial single pancake
regime, $B \sim \Delta t$, the sub-diffusive elastic line regime develops
and  $B \sim \Delta t^{\alpha}$. In the VL $\alpha = 1/2$, while below some
\textit{glass crossover} temperature to be determined below, $\alpha(T) <
1/2$. The scaling properties of $B(t,t_w)$ will be analyzed in Section
\ref{sec:scaling}. The same behavior is observed using $\gamma = 10^{-4}$.

\subsection{$\chi (t,t_w)$}

\begin{figure}[!tbp]
\includegraphics[angle=-90,width=\linewidth,clip=true]{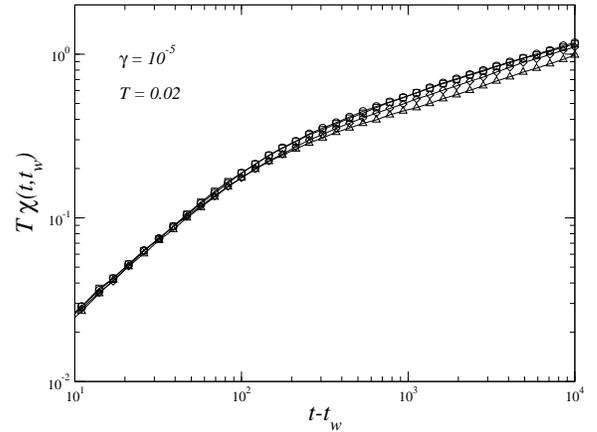}
\caption{\label{f:Xdt_400} Integrated response $T\chi(t,t_w)$, where
we include the temperature prefactor ($k_B=1$). The data correspond to
$\gamma = 10^{-5}$ and $T=0.02$.}
\end{figure}

In Fig.~\ref{f:Xdt_400} the dependence of  the response function $\chi(t,t_w)$ 
with $t$ and $t_w$
is shown for $\gamma = 10^{-5}$ and $T=0.02 < T_g$. There is also an aging
behavior, but not as severe as that of $B(t,t_w)$ for the same
parameters (Fig.~\ref{f:Bdt_b_418-410-407-403-400}, lower panel). This
fact is related to the type of scaling associated to the aging regime,
which will be analyzed in Sec.~\ref{sec:scaling}.

\subsection{Comparison with aging of the single elastic line}\label{sec:aging-single}

\begin{figure}[!tbp]
\includegraphics[angle=0,width=\linewidth,clip=true]{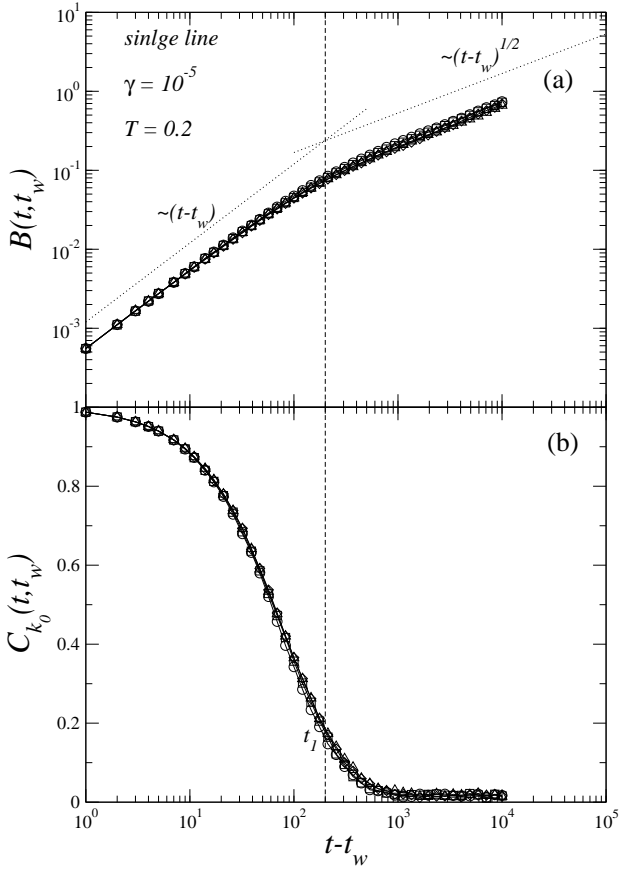}
\caption{\label{f:Bdt-Cdt_398} Stationary dynamics in the high-$T$ regime
($T=0.2$) of the single line, {\it i.e.} without in-plane interactions. (a)
$B(t,t_w)$ and (b) $C_{k_0}(t,t_w)$ for $\gamma=10^{-5}$ and different waiting
times. The data for $t_w=10^3,\,10^4,\,10^5$, and $10^6$ collapse on a master 
curve.}
\end{figure}

\begin{figure}[!tbp]
\includegraphics[angle=0,width=\linewidth,clip=true]{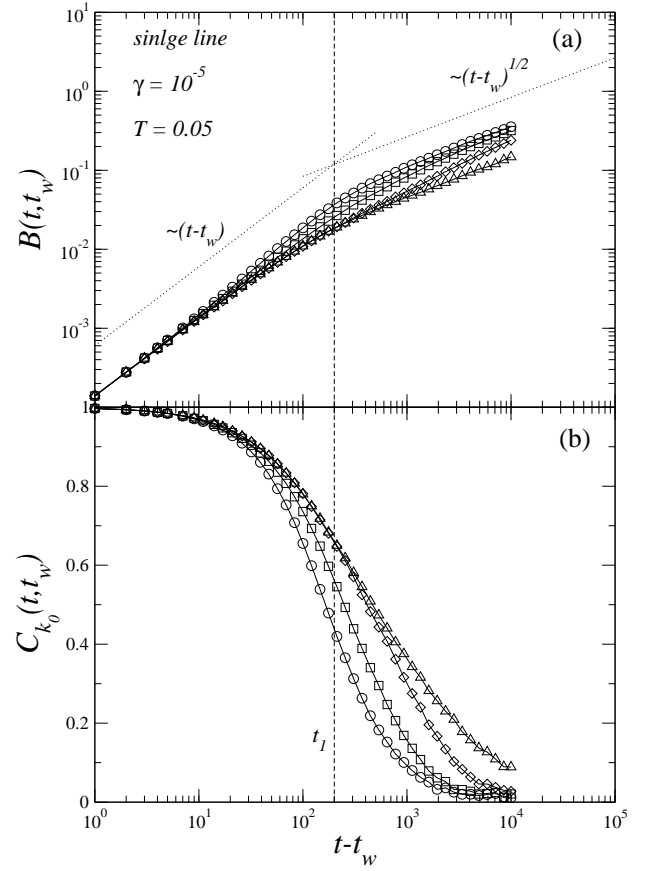}
\caption{\label{f:Bdt-Cdt_383} Aging in the low-$T$ regime ($T=0.05$) of 
the single line case, {\it i.e.} without in-plane interactions. 
(a) $B(t,t_w)$ and
(b) $C_{k_0}(t,t_w)$ for $\gamma=10^{-5}$ and different waiting times:
$t_w=10^3,\,10^4,\,10^5$ and $10^6$ from left to right.}
\end{figure}

We present here results for the out-of-equilibrium dynamics of single
elastic lines with disorder, {\it i.e.} without the in-plane
vortex-vortex interaction. In Figs.~\ref{f:Bdt-Cdt_398} and
\ref{f:Bdt-Cdt_383} we present the evolution with time of $B(t,t_w)$
and $C(t,t_w)$ for the high and low temperature regimes, $T=0.2$ and
$T=0.05$, respectively. At high temperature the dynamics is stationary,
not showing any $t_w$-dependence. At low temperature the system ages,
as observed in the $t_w$-dependence of both correlations in
Fig.~\ref{f:Bdt-Cdt_383}. These features were also observed in the
related model of the directed polymer in random media,~\cite{yosh98,yosh-unp} 
and are qualitatively the same as those
presented in Sect.~\ref{sec:aging-temp} for the full interacting VG. 
This suggests that the out-of-equilibrium
dynamics of the interacting vortex problem is dominated by the relaxation
of the elastic line. We return to this important issue
when analyzing the FDT violation in the next Section.

\section{Aging scaling and FDT violation}\label{sec:scaling}

\begin{figure}[!tbp]
\includegraphics[angle=0,width=\linewidth,clip=true]{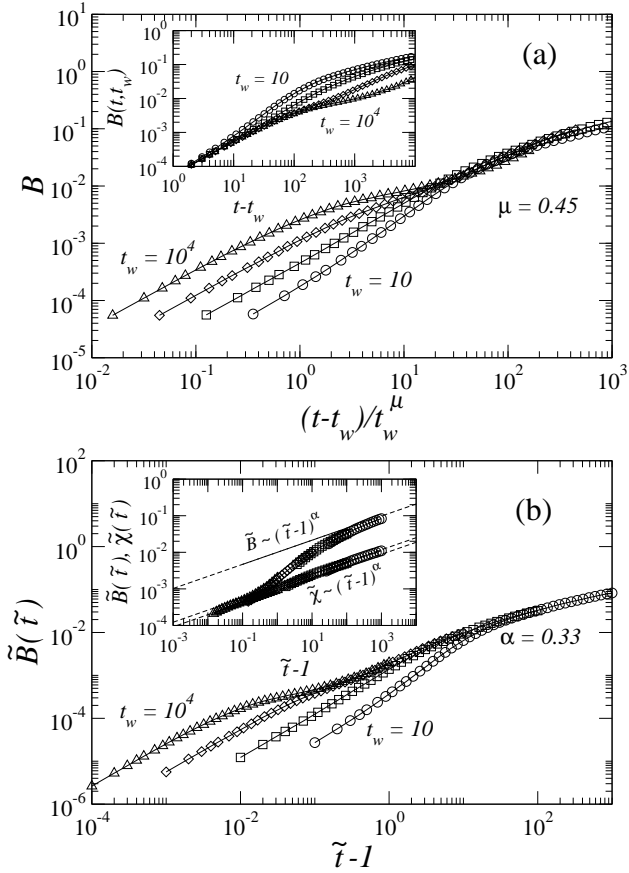}
\caption{\label{f:BgXgdt_400} (a) Test of a sub-aging additive
scaling for the pancake MSD for $\gamma = 10^{-5}$ and $T=0.02$.
The bare data are included in the inset. (b) Multiplicative
scaling for the same data as in (a). The values of the scaling
exponents $\mu$ and $\alpha$ are quoted. A better data collapse is observed in
the last case, including the short time regime. In the inset of
(b), the scaled MSD and the integrated response function are
shown. Different waiting times collapse on the same master
curve. The violation of the FDT at long scaled times $\widetilde t
\gg 1$ is clear.}
\end{figure}

\begin{figure}[!tbp]
\includegraphics[angle=-90,width=\linewidth,clip=true]{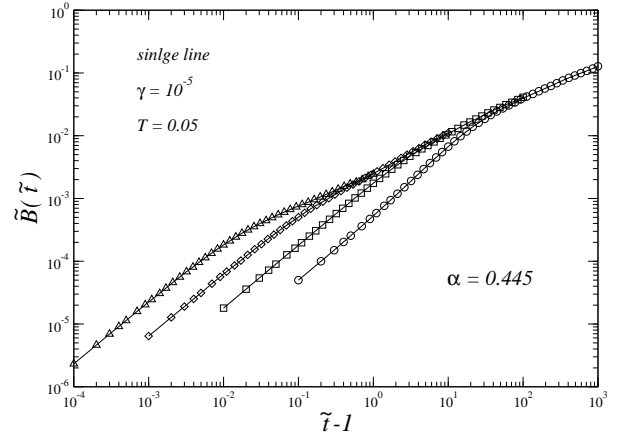}
\caption{\label{f:Bgdt_383} A single elastic line in a disordered environment:
multiplicative scaling of $B(t,t_w)$ for the data shown in
Fig~\ref{f:Bdt-Cdt_383}(a). $\gamma = 10^{-5}$ and $T=0.05$. The value of the
scaling exponent $\alpha$ is given in the key. Data for different waiting times
collapse, except for the stationary normal diffusion regime at very short
time-differences.}
\end{figure}

\subsection{Multiplicative aging}
\label{multip}

In this Section we study the two-times scaling of $B(t,t_w)$ and
$\chi(t,t_w)$ in the aging regime as well as the violation of the FDT
and the definition of the effective temperature.

In Fig.~\ref{f:BgXgdt_400} we show different scaling scenarii for $B(t,t_w)$
at $T=0.02$. On one hand, in Fig.~\ref{f:BgXgdt_400}(a) an additive sub-aging scaling is
put to the test by plotting $B(t,t_w)$ vs. $(t-t_w)/t_w^\mu$, while 
the unscaled $B(t,t_w)$ data is shown in the inset. On the other hand, in
Fig.~\ref{f:BgXgdt_400}(b) the multiplicative scaling of Eq.~(\ref{e:ag-mult}) is used
by plotting $\tilde B = t_w^\alpha\cdot B(t,t_w)$ vs. $(t-t_w)/t_w$. Although the
sub-aging scaling with a low $\mu$ exponent seems appropriate for long
time-differences we prefer the multiplicative scaling for at least two
important reasons.  First, the bare data shown in the 
inset to panel (a) do not show a
plateau at a  constant $B$ nor any clear trend to develop it at longer
$t_w$'s. This tends to disqualify the additive scaling for which a plateau is
expected. Second, the
multiplicative scaling allows for a better data collapse that includes the
short time-difference sub-diffusive regime, only leaving apart the very short
time-difference regime with normal diffusion. Thus, we find that our data is
very satisfactorily described within a \textit{multiplicative aging scaling}
scenario. This scaling is similar to the one proposed for the low temperature
behavior of the directed polymer in random media \cite{yosh98} 
and Sinai's diffusion.~\cite{Lale,Filemo,Lemofi}
Following Yoshino~\cite{yosh98} we tried the scaling form
\begin{subequations}
\begin{equation}
B(t,t_w)=t_w^{\alpha}\,\tilde B(\tilde t)
\label{eq:scaling-B}
\end{equation}
\begin{equation}
2k_B T\chi(t,t_w)=t_w^{\alpha}\,\tilde \chi(\tilde t),
\label{eq:scaling-chi}
\end{equation}
\end{subequations}
with $\tilde t=t/t_w$
and the scaled variables $\tilde B$ and $\tilde \chi$ given by
\begin{subequations}
\label{e:scalBX}
\begin{equation}
\label{e:tildeB}
\tilde B(\tilde t) = \left\{
\begin{array}{ll}
c_1(T)(\tilde t-1)^{\alpha(T)}
&
\qquad \;\;\;\;
\tilde t\ll1 \; , \\
c_2(T)(\tilde t-1)^{\alpha(T)}
&
\qquad \;\;\;\;
\tilde t\gg1 \; , \\
\end{array}
\right.
\\
\end{equation}
\begin{equation}
\label{e:tildechi}
\tilde \chi(\tilde t) = \left\{
\begin{array}{ll}
c_1(T)(\tilde t-1)^{\alpha(T)}
&
\;\;\;\;
\tilde t\ll1 \; ,  \\
y(T)c_2(T)(\tilde t-1)^{\alpha(T)}
&
\;\;\;\;
\tilde t\gg1 \; , \\
\end{array}\right.
\end{equation}
\end{subequations}
where $c_1$, $c_2$ and $y$ are temperature-dependent coefficients (see
the discussion in Sect.~\ref{log-power}). We assumed a linear
dependence of $\tilde \chi$ with $\tilde B$, but a non-trivial
function $\tilde \chi(\tilde B)$, as found in the scalar field or
Edwards-Wilkinson equation in $D=2$ and the spin-wave approximation to
the 2D XY model,~\cite{Cukupa,bert01} cannot be discarded.

In the inset to Fig.~\ref{f:BgXgdt_400}(b) the scaled $\tilde B$ and
$\tilde \chi$ at $T=0.02$ are shown. The data for different $t_w$ fall
on two master curves. $\tilde B$ and $\tilde \chi$ coincide for
$\tilde t \ll 1$, which means that FDT holds, while for longer times
$\tilde t\gg1$, the value $y(T)<1$ signals a violation of FDT, which
we will analyze in Sec.~\ref{vio-FDT}.

In Fig.~\ref{f:Bgdt_383} the multiplicative scaling for the single
line without in-plane interactions is shown. The curves correspond to
$\gamma = 10^{-5}$, $T=0.05$ and different waiting times. Besides
showing that this type of scaling is appropriate, this figure
emphasizes that the only regime not satisfying the scaling is the very
short stationary diffusion regime, as suggested from the general
picture in Fig.~\ref{f:sketch1}.

\subsection{Growing length}
\label{log-power}

Scaling arguments suggest that the averaged dynamics of an elastic
line in a random environment should be determined by a single growing
length scale, that separates equilibration at short length-scales from
non-equilibrium at long length-scales. This is also,
essentially, the picture that dictates dynamic scaling in coarsening
systems.

The multiplicative scaling of Eqs.(\ref{eq:scaling-B}) and
(\ref{eq:scaling-chi}) implictly assume a growing length with a
power-law dependence in time.  However, the analytic determination of
such a time-dependent length is very hard; it is known in a small
number of problems such as ferromagnetic domain growth in clean
systems with conserved and non-conserved order
parameter.~\cite{bray} But even these presumably simple cases can be
plagued with pre-asymptotic regimes, as shown for instance by
Krz\c{a}ka{\l}a for the 2D Ising model with Kawasaki dynamics.~\cite{Florent}

In the case of an elastic string in a random environment a
phenomenological creep argument complemented with the assumption that
typical barriers scale as $L^{\theta}$ leads to
\begin{equation}
L(t) \sim L_c \left[ \frac{T}{U_c} 
\ln\left(\frac{t}{t_0}\right)\right]^{1/\theta}
\label{e:Llog}
\end{equation}
{\it asymptotically}, {\it i.e.} 
in the limit of very long times and very large
scales. $L_c$ is the Larkin length, $U_c$ its corresponding energy
scale and $t_0$ a microscopic time scale. The exponent $\theta$ is
usually further assumed to take the same value as for the free-energy
fluctuations, $\theta=1/3$. Alternatively, a similar argument with
logarithmically growing barriers  imply a power law
growth of the typical length scale $L(t)$.

Yoshino \cite{yosh, yosh-unp} and Kolton {\it et al}~\cite{kolt05b}
studied the growing length-scale numerically in a lattice and a continuous 
model of a single line in a random environment, respectively. Both studies 
show that $L(t)$ crosses over from a power-law to a logarithmic growth
at a typical time-scale $t^*$ associated to a typical length scale $L^*$. 
For instance, Kolton {\it et al} found $t^*\sim 10^4$ and $L^*\sim 80$
using lines with total lengths $L=256$ and $L=512$. Similar values were 
found by Yoshino. 

In our study of vortices {\it in interaction} we were forced to use
rather short total lengths, $L=50$. Hence, we may suppose our dynamics
falls in the first regime in which $L(t)$ grows as a power
law. Indeed, all our data can be satisfactorily scaled in a manner
that is consistent with such a power law. We may expect, clearly, 
that the crossover in the growing length will induce a 
change in the scaling forms (\ref{eq:scaling-B})  and (\ref{eq:scaling-chi})
possibly to a more general form
\begin{eqnarray}
B(t,t_w) &\sim& L^\zeta(t_w) \; \widetilde{B}_{ag}\left(\frac{L(t)}{L(t_w)}\right),
\\
\chi(t,t_w) &\sim& L^{\overline\zeta}(t_w) \; \widetilde{\chi}_{ag}
\left(\frac{L(t)}{L(t_w)}\right). 
\end{eqnarray}
For times $t < t ^*$, with a behavior $L(t) \sim t^{\alpha/\zeta}$ 
and $\overline\zeta=\zeta$ one recovers
Eqs.~(\ref{eq:scaling-B})  and (\ref{eq:scaling-chi}). For longer times
a growing $L(t)$ as in Eq.~(\ref{e:Llog}) is also possible. 
Testing the latest time-regime, however, goes beyond the scope of this study.  
The value of the exponent $\overline\zeta$ is not completely clear 
yet. Our numerical data support $\overline\zeta=\zeta$ but at longer
times one might notice a deviation. 

\begin{figure}[!tbp]
\includegraphics[angle=-90,width=\linewidth,clip=true]{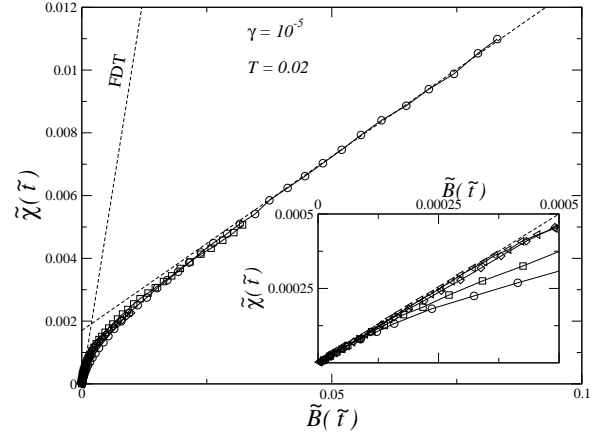}
\caption{\label{f:XgBg_400} Parametric plot $\tilde \chi(\tilde B)$ in the VG
(same data as in Fig.~\ref{f:BgXgdt_400}). FDT holds at short rescaled-times
while a violation of FDT with $y < 1$ appears at longer rescaled-times. The
inset shows the short rescaled-time FDT regime for different $t_w$ values.}
\end{figure}

\begin{figure}[!tbp]
\includegraphics[angle=-90,width=\linewidth,clip=true]{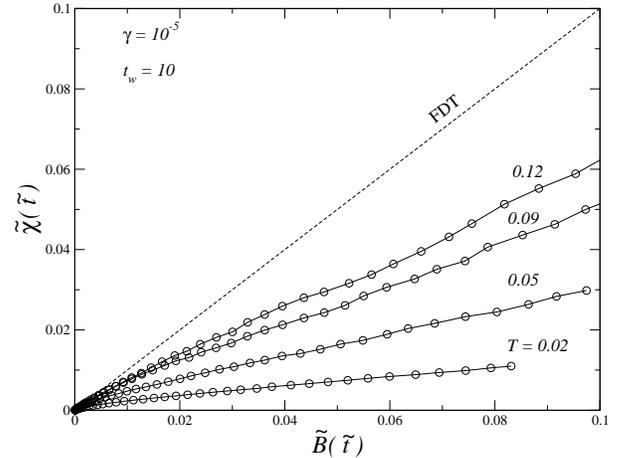}
\caption{\label{f:XgBg_400-403-407-410} Parametric plot $\tilde \chi(\tilde
B)$ in the VG at different temperatures given in the figure.}
\end{figure}

\subsection{Violation of FDT and effective temperature}
\label{vio-FDT}

In order to test the violation of FDT, 
we use our results of Sec.\ref{multip}  scaled with 
Eqs. (\ref{eq:scaling-B})  and (\ref{eq:scaling-chi}).
The parameter $y(T)$ in  Eqs.~(\ref{eq:scaling-B})  and (\ref{eq:scaling-chi})
 measures the modification of the FDT,  $2k_B T\chi(t,t_w)=
y(T)B(t,t_w)$ (or $\tilde \chi=y(T)\tilde B$), and an effective
temperature \cite{peliti} can then  defined by $T_{\rm eff}=T/y$ (see
Eq.~\ref{e:BXfdt}).
 
\begin{figure}[!tbp]
\includegraphics[angle=0,width=\linewidth,clip=true]{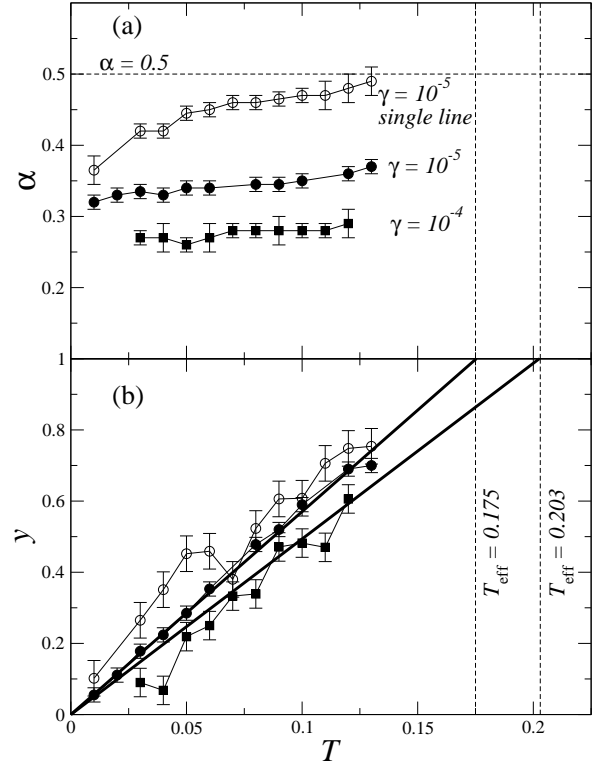}
\caption{\label{f:ay} (a) The scaling exponent $\alpha(T)$ and (b) the $y(T)$
parameter measuring the FDT violation in the VG with two disorder intensities,
$\gamma=10^{-5}$ and $\gamma=10^{-4}$. Data for a single elastic line
in a random environment with $\gamma=10^{-5}$ are also shown. 
The effective temperature
$T_{\rm eff}$ obtained from linear fits $y(T)=T/T_{\rm eff}$ is given for
the cases $\gamma=10^{-5}$ and $\gamma=10^{-4}$.}
\end{figure}

In Fig.~\ref{f:XgBg_400} we show a parametric plot of the scaled
variables, $\tilde \chi (\tilde t)$ vs. $\tilde B(\tilde t)$,
for a temperature $T < T_g$. 
If FDT is violated then the slope of the parametric plot
should be different from unity. The parameters in
Fig.~\ref{f:XgBg_400} are $\gamma = 10^{-5}$ and $T=0.02$, and
different waiting times are shown. At very short rescaled times $\tilde t$
all curves fall
onto the FDT line, while for longer times the curves asymptotically
change to a different slope $y(T)$. In the inset, we show the short
time regime for different $t_w$ values. Since the data with lower
waiting time, $t_w=10$, is deeper in the out-of-equilibrium regime, we
used the slope of $\tilde \chi(\tilde B)$ with $t_w=10$ to compute the
$y(T)$ parameter signaling the FDT violation. In
Fig.~\ref{f:XgBg_400-403-407-410} we show $\tilde \chi(\tilde B)$ with
$t_w=10$ at different temperatures; it is clear in the figure that the $y(T)$
parameter decreases with decreasing temperature. These and similar
curves are used to compute the $y(T)$ values showed in
Fig.~\ref{f:ay}(b). Exactly the same behavior is obtained for the
single elastic line (not shown).

In order to investigate the temperature dependence of different quantities in
the low temperature regime, we show the temperature variation of the scaling
exponent $\alpha(T)$ and the $y(T)$ parameter in Fig.~\ref{f:ay}. We use two
values of the disorder intensity, $\gamma=10^{-5}$ and $\gamma=10^{-4}$. Also
shown are data for $\gamma=10^{-5}$ but for the case without in-plane
interactions, quoted as \textit{single line}. The values of the exponent and
FDT violation factor at high temperature are $\alpha = 1/2$, corresponding to
the single elastic line without disorder limit (or VL phase), and $y=1$,
corresponding to the equilibrium FDT. In Fig.~\ref{f:ay}(a) we see that
$\alpha(T)$ depends weakly on $T$ and that it decreases with increasing
disorder strength within the glassy regime. It is also observed that
the in-plane interactions tend to decrease the value of $\alpha$. The parameter
$y(T)$ measuring the violation of FDT rapidly increases with increasing
temperature. In Fig.~\ref{f:ay}(b) we show that $y(T)=T/T_{\rm eff}$ is well
described by a linear form, implying an effective temperature $T_{\rm eff}$
that is temperature-independent. However, a more complicated
temperature-dependent effective temperature cannot be ruled out. We do not
show the linear fit corresponding to $\gamma=10^{-5}$ without in-plane
interactions. From the linear fit of $y(T)$ one obtains the effective
temperature values quoted in the figure, $T_{\rm eff} = 0.175$ and $T_{\rm
eff} = 0.203$ for $\gamma=10^{-5}$ and $\gamma=10^{-4}$,
respectively.

\begin{figure}[!tbp]
\includegraphics[angle=-90,width=\linewidth,clip=true]{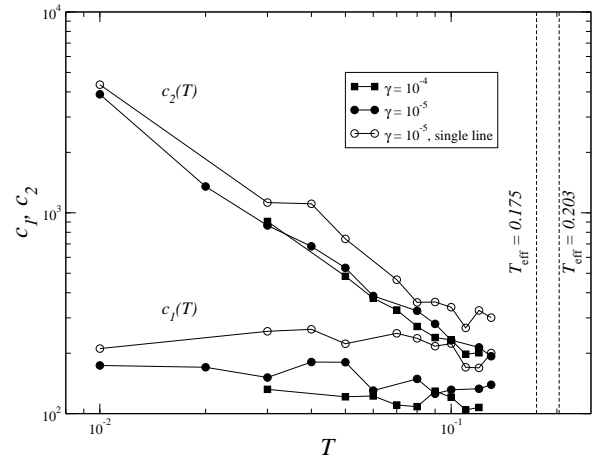}
\caption{\label{f:cdT} Evolution of the coefficients $c_1(T)$ and $c_2(T)$ 
with temperature for the same parameters as in Fig.~\ref{f:ay}. They tend to
the same value with increasing temperature, as suggested by the scaling form
(\ref{e:scalBX}).}
\end{figure}

Another feature of the scaling form (\ref{e:scalBX}) is the temperature
dependence of the $c_1(T)$ and $c_2(T)$ coefficients. These coefficients
measure the separation between the quasi-equilibrium regime and the aging
regime, and hence they should converge to the same value in the equilibrium
high-temperature regime. In Fig.~\ref{f:cdT} these functions are shown for the
same parameters used in Fig.~\ref{f:ay}. The $T_{\rm eff}$ values are also
given in the figure. For each set of parameters $c_1(T)$
and $c_2(T)$ are closer to each other when approaching $T_{\rm eff}$ and the
high-temperature regime. This is another confirmation of the scaling form
(\ref{e:scalBX}).

 Noticeably, the obtained value of $T_{\rm eff} = 0.175$ is very close
to the crossover temperature $T_g\approx0.18$ 
below which a dynamic arrest is observed and the system can not be equilibrated,
as found in Sec.\ref{sec:diff}. A similar result is observed in
structural glasses: $T_{\rm eff}\approx T_g$ (as in a random energy model
scenario~\cite{parisi}).

\section{Discussion}\label{sec:conc}

In this paper we studied numerically the dynamics of a high-temperature
superconductor model. We briefly analyzed the stationary dynamics in the VL
and we focused on the out-of-equilibrium dynamics of the VG. Our aim was to
learn about the nature of the VG and its dynamical properties from the
relaxation of different two-times correlation functions.  During this study a
systematic comparison with the dynamics of other glassy
systems was performed.

A key feature in problems involving elastic lines, as the one we
treated here, that should be highlighted is the relevance of finite
size effects. By comparing the evolution of the roughness and dynamic
wandering in the vortex system to the ones in the EW equation for
growing interfaces \cite{barabasi} we identified three dynamic
regimes present {\it at all} interesting 
temperatures: (\textit{i}) a very short time-difference normal diffusion
regime without effective elastic interactions; (\textit{ii}) a
sub-diffusion intermediate time-difference regime characterized by a
growing longitudinal correlation length; (\textit{iii}) a long
time-difference regime where the correlation length has reached the
system size and normal diffusion simply reflects the center of mass
diffusion. In order to separate aging effects from finite size effects
we constrained the remaining study to regimes (\textit{i}) and
(\textit{ii}), {\it i.e.} before roughness saturation.

Strikingly, the behavior of the density-density correlation function
$C_k(t,t_w)$ resembles in many aspects the one observed in simulations of
Lennard-Jones glass formers.~\cite{kob00} In particular, the wave-vector
dependence of the correlation follows the same general trend. The main
difference is that the density-density correlation does not develop a well
defined plateau for long waiting times in the VG, while it seems to do in 
glass-forming liquids. This is not due to the
use of extremely short waiting times as could be expected in an additive aging
scaling, but it is in the core of the multiplicative aging scaling we found in
the present simulations. 

We showed that the two-times evolution of the pancake 
mean-squared-displacement $B(t,t_w)$ is very well described by the multiplicative
scaling diffusive-aging scenario earlier proposed in the numerical
study of the out-of-equilibrium dynamics of the directed polymer in
random media model,~\cite{yosh98} found analytically in the massless
scalar field in $D=1$ (related to the Edwards-Wilkinson surface), and
obtained numerically and analytically in Sinai
diffusion.~\cite{Lale,Filemo,Lemofi} It is interesting to stress that
the model of a $D$-dimensional manifold embedded in an $N$-dimensional
space under the effect of a combined disordered and harmonically
confining potential -- a toy model that is usually used to model
vortex systems -- fails to capture the multiplicative scaling in the
large $N$ limit.~\cite{Cule,Cukule} This model has, instead, a well
defined transition at a finite temperature below which correlations
and responses scale in an additive way with a stable plateau.

We studied the violations of the FDT out of equilibrium.  We
constructed the parametric plot of linear integrated response against
displacement after having eliminated the power of the waiting-time,
$t_w^\alpha$, that appears multiplying these quantities.  The
resulting plot is linear, at least within the accuracy of out data,
and it allowed us to identify an effective temperature as the slope of
the linear plot. The same behavior was found in the lattice model of
the directed polymer in a random environment.~\cite{abarrat,yosh} The
relation between integrated response and displacement, once the
factors $t_w^\alpha$ have been taken into account, might involve
though a non-trivial function of the displacement itself, not visible
within the accuracy of our data, as has been found, for instance, in
the 2D scalar field and XY model.~\cite{Cukupa,bert01} 
The FDT violation in the `mean-field' model of a
relaxing $D$-dimensional manifold in an infinite-dimensional embedding
space~\cite{Cule,Cukule} with a short-ranged disorder potential has
this type of violation of FDT though it is not necessary in this case
to eliminate additional $t_w^{\pm \alpha}$ factors.

We performed the analysis of displacement, correlations and responses
at several low temperatures and we analysed the temperature dependence
of the exponent $\alpha(T)$ as well as the coefficients $c_1(T)$ and
$c_2(T)$ that characterize the scaling function.  We also followed the
evolution with temperature of the parameter signaling the FDT
violation, $y(T)$.  We considered two intensities of the pinning
disorder and we compared the data to the {\it a priori} simpler case
of a single line without in-plane interactions. All the parameters,
$y(T)$, $\alpha(T)$, $c_1(T)$ and $c_2(T)$ show a clear trend to reach
the high-temperature behavior: $y(T)=1$, $\alpha(T)=1/2$, and
$c_1(T)=c_2(T)$ when approaching the VL from the glassy regime.  From
these out-of-equilibrium measurements the crossover temperature was
found to be very similar to the value of the effective temperature,
$T_{\rm eff}$, obtained from the fit of the FDT violation at very long
time-differences to a linear form with $y(T)=T/T_{\rm eff}$.  This
result is similar to what happens in the random-energy model in
which the effective temperature takes the transition value in the full
low-temperature phase.

To further test the meaning of the crossover temperature and the effective
temperature $T_{\rm eff}$, we studied the dynamic arrest approaching the VG
from the VL. By fitting a mean-squared-displacement correlation, taking into
account sub-linear diffusion, we determined a relaxation time $t_r$. The
so-obtained relaxation time increases with temperature as $t_r \sim T^{-2}$,
at rather high temperatures, as expected in disordered free dynamics. However,
below a characteristic temperature a sudden further increase of the relaxation
time $t_r$ was found. We identified the crossover temperature between normal
relaxation and rapidly growing  relaxation with a glass temperature $T_g$. When
comparing with the effective temperature we found that both temperatures are
of the same order, a fact already observed in other glassy systems. This
suggest a freezing of the slow degrees of freedom of the system around the
crossover temperature.

One outstanding result of our simulations is that all the
out-of-equilibrium properties are mainly dictated by the elastic line
energy, {\it i.e.} they are all observable in the single line
without in-plane interactions. This means that for the parameters used
in our simulations the two relevant competing energies are the elastic
line and pinning energies. This fact points out that the relevant
dynamic properties are also observed in simplified models of elastic
lines in disordered media as directed polymers
\cite{yosh98,yosh-unp,bust-unp} or continuous models similar to that
studied here.~\cite{kolt05a,kolt05b} It is worth mentioning that this
cannot be the case at very high magnetic fields, {\it i.e.} very
high vortex density, since at some point the in-plane interaction
should become the relevant energy scale. In this condition strong
excluded volume effects, such as the ``cage effect'' in supercooled
liquids, should be appreciable. A simplified model in this direction
was considered by Pet\"aj\"a {\it et al.}, who considered elastic
lines with hard-core interactions in random environments.~\cite{peta04} 

From our findings several extensions could be envisaged. Concerning
the possible experimental observation of the out-of-equilibrium
dynamics of the VG phase, new experiments monitoring magnetic
relaxation \cite{papa99,papa02} would be most welcome.
To measure aging, one possibility is to perform transport
experiments following a protocol similar to the one used
by Ovadyahu and coworkers to study the electron glass regime
in Anderson insulators .~\cite{ovadyahu00,ovadyahu02}
Tests of FDT would require simultaneous measurements of 
noise (magnetic and/or voltage noise) and time dependent
response (relaxation of magnetization and/or resistivity);
similar tests have been performed in other systems 
.~\cite{grigera,bellon,herisson} Another
possibility would be to test the aging dynamics and the FDT violation
in vortex shear experiments.~\cite{lope99,broj01} It is well known
that in shear-thinning systems the time scale introduced by the shear
rate basically replaces the waiting time; at lower shear rates the
relaxation is longer, {\it i.e.} the system ``ages'' with decreasing
shear rate.~\cite{bert00,barr00,bert02a,bert02b} Still another
possibility is to measure effective temperatures using a tracer
embedded in the vortex systems. Finally, the out-of-equilibrium
dynamics of the center of mass diffusion could be experimentally
tested.

It would also be interesting to study the out-of-equilibrium 
relaxation of the BG and compare the outcome 
to the results obtained in this 
work for the VG. 

\acknowledgments

We thank C. Chamon, A. B. Kolton, T. Giamarchi, J. L. Iguain and H.
Yoshino for very useful discussions and H. Yoshino for access to his
unpublished results on the dynamics of a directed polymer in a random
environment.  We acknowledge financial support from SECYT-ECOS P.
A01E01 (SB, LFC DD); Conicet PIP2005-5596, CNEA P5-PID-93-7, ANPCYT
PICT99-03-06343, and ICTP grant NET-61 (SB and DD); Fundaci\'on
Antorchas (SB) and PICS 3172 (LFC).  SB thanks the LPTHE Jussieu
(France), SB and LFC the Universidad Nacional de Mar del Plata
(Argentina), and LFC the Newton Institute for Mathematical Sciences
(UK) for hospitality during the preparation of this work.  LFC is a
member of Institut Universitaire de France.

\bibliography{vortex-glass}

\end{document}